\documentclass[11pt]{article}
\usepackage{latexsym}
\usepackage[english]{babel}
\usepackage{amssymb,amsmath,amsfonts,amsthm}
\usepackage{dsfont,mdframed,mathrsfs,stmaryrd}
\usepackage[active]{srcltx}
\usepackage{graphicx,framed}
\usepackage[table]{xcolor}
\usepackage{scrtime}
\usepackage{cite}
\usepackage[normalem]{ulem}
\usepackage{mathtools}
\usepackage{comment,cancel}
\usepackage{units}
\usepackage{parskip}
\usepackage{braket}
\usepackage{hyperref}
\usepackage{float}
\usepackage[margin=1in]{geometry}
\usepackage[toc,page]{appendix}
\usepackage{multirow}
\usepackage{array}

\usepackage{tcolorbox} 
\tcbset{
  colback=white    
}

\usepackage{multirow}

\usepackage{soul}

\usepackage{tikz}

\newcolumntype{H}{>{\setbox0=\hbox\bgroup}c<{\egroup}@{}}

\hypersetup{
linktoc=all,
    citecolor=black,
    filecolor=black,
    linkcolor=black,
    urlcolor=black
}

\makeatletter

\@addtoreset{equation}{section}
\makeatother

\def\be{\begin{equation}}
\def\ee{\end{equation}}
\def\bea{\begin{eqnarray}}
\def\eea{\end{eqnarray}}
\def\nn{\nonumber}

\def\m{\mu}
\def\n{\nu}

\newcommand{\eq}[1]{(\ref{#1})}
\newcommand{\w}[1]{\\[0.#1cm]}

\DeclareMathOperator{\Tr}{Tr}
\DeclareMathOperator{\tr}{tr}

\newcommand{\bbR}{\mathbb{R}}
\newcommand{\bbZ}{\mathbb{Z}}
\newcommand{\SO}[1]{\mathrm{SO}(#1)}
\newcommand{\SU}[1]{\mathrm{SU}(#1)}
\newcommand{\Sp}[1]{\mathrm{Sp}(#1)}
\newcommand{\Uni}[1]{\mathrm{U}(#1)}

\definecolor{darkgreen}{rgb}{0.42, 0.56, 0.14}
\definecolor{darkred}{rgb}{0.65,0.15,0}

\allowdisplaybreaks

\def\bR{\mathbb{R}}
\def\bZ{\mathbb{Z}}
\def\cA{\mathcal{A}}
\def\cE{\mathcal{E}}
\def\Ext{\operatorname{Ext}}

\definecolor{wo}{rgb}{1.0, 0.85, 0.6}
\definecolor{wb}{rgb}{0.7, 0.9, 1.0}
\newcommand{\tikzmark}[2]{\tikz[remember picture,overlay]\node[anchor=base] (#1){#2};}

\def\sG{\mathsf{G}}

\def\tv{{\widetilde v}}

\begin{document}

\thispagestyle{empty}

\begin{flushright}\small
MI-HET-857   \\

\end{flushright}

\bigskip\bigskip

\vskip 10mm

\begin{center}

{\Large{\bf Global anomalies in $6D $ gauged supergravities
}}

\vskip 4mm

\end{center}

\vskip 6mm

\begin{center}

\normalsize

{\large K. Becker$^\star$, E. Sezgin$^\star$ and D. Tennyson$^\star$ } \\

{\large with a mathematical appendix by Y. Tachikawa$^\dagger$}

\vskip 8mm

$^\star$\,{\it George P. and Cynthia W. Mitchell Institute \\for Fundamental
Physics and Astronomy \\
Texas A\&M University, College Station, TX 77843-4242, USA}

$^\dagger$\,{\it Kavli Institute for the Physics and Mathematics of the Universe (WPI),\\
University of Tokyo, Kashiwa, Chiba 277-8583, Japan}

\end{center}

\vskip1.5cm

\begin{center} {\bf Abstract } \end{center}

There exists a rare class of R-symmetry gauged $N=(1,0)$ supergravities in six dimensions with gauge group $G\times U(1)_R$, where $G$ is semisimple with rank greater than one, and the number of tensor multiplets $n_T=1$, which are free from all local anomalies. We find new members of this family, in which $G$ contains up to four factors. We study the global anomalies of these models in a framework in which the Dirac quantization of the anomaly coefficients and the well-definedness of the Green-Schwarz anomaly counterterm in generic backgrounds play key roles, and we apply the anomaly freedom criteria that takes this into account as formulated by Monnier and Moore in \cite{Monnier:2018nfs}. To this end we use the result derived by Yuji Tachikawa in the appendix which states that the spin cobordism group $\Omega_7^{\rm Spin}(BG)$ vanishes for $G=\sG_1\times \cdots \times \sG_n$ where $\sG_i$ is $U(1)$ or any simple simply-connected non-Abelian compact Lie group. We also require correct sign for the vector field kinetic terms, and positive Gauss-Bonnet term. We find that among the models considered here only one of them fails to satisfy all the stated criteria.  We also find that, in general, the requirement that the anomaly coefficients defined by the factorized anomaly polynomial are elements of a unimodular charge lattice imposes a constraint on the number of vector multiplets given by $n_{V}=8\ {\rm mod}\ 12$ for $\Uni{1}_{R}$, and $n_{V} = 12$ or $96$ for $\Sp{1}_{R}$ gauged models. 

\newpage

\tableofcontents

\section{Introduction}

The premise of the Swampland program initiated in \cite{Vafa:2005ui} is to study matter coupled to (super)gravity theories that satisfy criteria based on universal physical considerations and thereby determining if they can in principle be candidates for a UV completion without having to be necessarily related to string theory. In this endeavor, a special class of $R$-symmetry gauged $N=(1,0)$ supergravities in six dimensions provide a very fruitful ground. The salient features of these theories include the fact that the gravitini are charged under the $R$-symmetry group, which is $Sp(1)_R$ or $U(1)_R$, and the potential is non-vanishing even upon setting all scalar fields equal to zero,  yielding a cosmological constant. Those with single tensor multiplet coupling that admit Green-Schwarz anomaly cancellation mechanism are rare if one seeks semisimple gauge groups whose factors other than the R-symmetry group have rank greater than one. So far, 9 such models have been found \cite{Avramis:2005hc,Randjbar-Daemi:1985tdc,Avramis:2005qt,Becker:2023zyb}. Interestingly in all cases the hyperscalars are only in non-trivial irreps of the gauge group. Another remarkable aspect of these models is that they are not known to arise from string/M/F theory by any mechanism known to date.

In this paper, we have two goals. First, we will present thirteen new local anomaly-free models of the kind described above. The details of these models and the assumptions made in the search will be discussed in Sec.~2. Second, in the spirit of the Swampland program mentioned above, we shall study the consistency conditions of the rare class of R-symmetry gauged models, which now has 24 members, from the requirement of the absence of global anomalies. We use the terminology of ``global anomalies'' to refer to more than the large gauge transformations, but also to Dirac quantization conditions that arise in the attendant dyonic string solutions, as well as the well-definedness of the Green-Schwarz anomaly counterterms, 
as has been formulated by Monnier and Moore in \cite{Monnier:2018nfs}, building up on 
\cite{Kumar:2010ru,Seiberg:2011dr,Monnier:2017oqd}. For a summary and further references, see \cite{Monnier:2018cfa}. 
These criteria were applied in \cite{Pang:2020rir} to previously known 3 models, and only one of them was shown to satisfy them. One of these models failed because it involved a non-simply connected gauge group. Here we shall consider only simply connected gauge groups and we will see that two of those three models, which have the gauge groups $E_6\times E_7\times U(1)_R$ and $F_4\times {\rm Sp}(9)\times U(1)_R$, are free from global anomalies. We shall also show that the remaining 21 models satisfy all the criteria of \cite{Monnier:2018cfa} for global anomaly freedom as well. 
These criteria include the vanishing of the spin bordism group $\Omega_7^{\rm Spin}(BG)$ for $G$ which is a product of various classical and exceptional Lie groups. This condition arises from the requirement that the Green-Schwarz anomaly counterterm is globally well defined in generic backgrounds as explained in \cite{Monnier:2018nfs}. The vanishing of $\Omega_7^{\rm Spin}(BG)$ was shown in \cite{Monnier:2018nfs} for $G$ which is any product of $U(1), SU(n)$ and $Sp(n)$. Here, this result is extended in the appendix provided by Yuji Tachikawa to 
any products of $U(1)$ and simple simply-connected non-Abelian compact Lie groups.

In analyzing the consequences of the global anomaly cancellation in the $R$-symmetry gauged theory for $n_T=1$, we shall find a constraint on the total number of vector fields, $n_V$, given by 
\begin{equation}
\Uni{1}_R:\ \quad n_V \equiv 8 \mod 12\ ,\qquad  \Sp{1}_R:\quad n_V = 12\ \mbox{or} \ 96
\label{ncs}
\end{equation}
In particular, the requirement that the charge lattice should be unimodular, as derived by Seiberg and Taylor in \cite{Seiberg:2011dr}, leads to the above conditions on the number of vector-multiplets $n_V$.
It is remarkable that the dimension of the gauge group in the case of $Sp(1)_R$ gauging is greatly restricted. Furthermore, the derivation of these constraints crucially depends on the presence of $R$-symmetry gauging, and indeed they are absent for the $R$-symmetry ungauged models. 

This paper is organized as follows. In Sec.~2, we briefly review the structure of the known local anomaly-free $R$-symmetry gauged $6D$ supergravities, and their anomaly polynomials. In this section we also present new additions to this family. In Sec.~\ref{sec:global_anomalies} we review the criteria for global anomaly freedom as formulated in\cite{Monnier:2018cfa}, and apply these criteria to the models described in Sec.~2. We also derive the constraint \eqref{ncs} on the number of vector fields. Sec.~5 contains a brief summary and further comments. The appendix, provided by Yuji Tachikawa, provides the computation of the spin bordism group $\Omega_7^{\rm Spin} (BG)$ for $G$ mentioned above. 

In the remainder of this paper, for simplicity, we shall often refer to the $R$-symmetry gauged supergravities simply as ``gauged supergravities". These are not to be confused with supergravities coupled to Yang-Mills in which the gauge group does not have an $R$-symmetry group factor, which are sometimes referred to as ``gauge supergravities" in the literature.

\section{Anomaly-free gauged \texorpdfstring{$N=(1,0)$}{N=(1,0)}, 6D supergravities: Old, recent and new}\label{sec:gauged_review}

The $N=(1,0)$ super-Poincar\'e algebra in $6D$ admits the following multiplets
%
\be
\underbrace{ \left(e_\m^m, \psi_{\m A}^+, B_{\m\n}^{+} \right)}_{\rm graviton}\ ,\qquad \underbrace{\left(B_{\m\n}^{-}, \chi_A^-,\varphi  \right)}_{\rm tensor}\ ,\qquad \underbrace{\left(A_\m, \lambda_A^+\right)}_{\rm vector}\ ,\qquad \underbrace{\left(4\phi,\psi^{-} \right)}_{\rm hyper}\ .
\ee
The two-form potentials, $B_{\m\n}^{\pm}$, have (anti-)self-dual field strengths. The spinors are symplectic Majorana-Weyl, $A=1,2$ labels doublet of $R$ symmetry group, and  chiralities of the fermions are denoted by $\pm$. The spacetime signature is $(-+++++)$ and the chirality gamma matrix is $\gamma_7=\gamma_0\gamma_1...\gamma_5$.  We consider $N=(1,0), 6D$ supergravity with $U(1)_R$ $R$-symmetry gauged, and coupled to a single tensor multiplet (thus $n_T=1$), $n_V$ vector and $n_H$ hypermultiplets such that the total gauge symmetry group is 
\be
G=G_1\times\cdots \times G_n\times U(1)_R\ ,
\label{g1}
\ee
where $G_1\times\cdots \times G_n$ is semi-simple. Denoting by $n_R^i$ the number of times the representation $R$ of $G_i$ is carried by the hyperfermions, and by $n_{RS}^{ij}$ the number of times they carry the representations $R$ and $S$ of the group factors $G_i$ and $G_j\,  (i<j)$ simultaneously, the sum of all the anomaly polynomials in Lorentzian signature spacetime 
\footnote{The one loop anomaly arising from the variation of the one loop effective action in Minkowski spacetime is $\delta \Gamma= \int I_6^1$, where $I_6^1$ is obtained by descent equations $I_8=dI_7^1$ and $\delta I_7^0=d I_6^1$, and $I_8= 2\pi \times \mbox{(index density)}$. The index density for different spins given in \cite{Alvarez-Gaume:1984zlq} is in Euclidean space, and in passing over to Minkowski spacetime, we note that $\gamma_7^E= -\gamma_7$.} 
\begin{align}
I_8 &=  \frac{2\pi}{(16\pi^2)^2}\Bigg[ (\tr R^2)^2 +\frac16  \left( -20 + n_V \right) F_{R}^2 \tr R^2
\nn\w2
& -4  F_{R}^2  \sum_{i=1}^n a_i^{\rm adj} \tr F_i^2 +4 \sum_{i<j} \sum_{R,S} n^{ij}_{RS} \tr_R F_i^2 \tr_S F_j^2 
\nn\w2
& -\frac23  (4 +n_V) F_{R}^4  -\frac23  \sum_{i=1}^n \left( c^i_{\rm adj} -\sum_R n_R^i c_R^i \right) (\tr F_i^2)^2
\nn\w2
& +\frac16  \tr R^2 \sum_{i=1}^n \left( a^i_{\rm adj} -\sum_R n_R^i a_R^i \right) \tr F_i^2\Bigg]\ ,
\label{me1}
\end{align}
with
\begin{align}
\tr_R F_i^2 &= a_R^i\, \tr F_i^2\ ,
\nn\w2
\tr_R F_i^4 &= b_R^i \tr F_i^4 +  c_R^i \tr (F_i^2)^2 +d_R\, {\rm Pf}(F) \ ,
\nn\w2
n_V &= 1+ \sum_{i=1}^n {\rm dim}\, G_i \ ,
\label{gtr}
\end{align}
where $d_R$ is the Pfaffian invariant defined as
\be
{\rm Pf} (F) := \frac{1}{2^4 4!} \epsilon^{abcdefg} \tr \left( F_{ab} F_{cd} F_{ef} F_{gh} \right)\ , 
\ee
and it is relevant only for $D_4$. 
Similar relations define the constant coefficients $a_{\rm adj}^i, b^i_{\rm adj}, c_{\rm adj}^i$, and $\tr$ is the trace in the fundamental representations. For any representation $R$ of any Lie group of dimension $d$, rank $r$, fundamental and adjoint representations $F$ and $A$ respectively, the coefficients $a_R, b_R, c_R$, {\it with the exception of $D_4$},  are given by \cite{Okubo:1981td,Avramis:2005hc}
\footnote{The indices $l_2$ and $l_4$ are tabulated for a large number of representations of all Lie groups in \cite{McKayPateraBook}, and they can easily be computed by using the LieART package \cite{Feger:2019tvk} in Mathematica for any representation of any Lie group. Formulae for the modified 4th index $\ell_4^\prime$ has been computed in \cite{Okubo:1981td} also for arbitrary representations of all Lie groups.}
\begin{align}
    a_{R} &= \frac{l_{2}(R)}{l_{2}(F)} \, , \\
    b_{R} &= \frac{l'_{4}(R)}{l'_{4}(F)} \, , \qquad l'_{4}(R) = l_{4}(R) - \frac{r+2}{r} \frac{d}{d+2} \left( \frac{l_{2}(R)}{\dim R} - \frac{1}{6}\frac{l_{2}(A)}{d} \right) l_{2}(R) \, , \\
    c_{R} &= \frac{3}{2+d} \left( a_{R}^{2}\left(\frac{d}{\dim R} - \frac{1}{6} \frac{l_{2}(A)}{l_{2}(R)}\right) - b_{R}\left( \frac{d}{\dim F} - \frac{1}{6} \frac{l_{2}(A)}{l_{2}(F)}\right) \right)\ ,
\end{align}
where the $2n^{\text{th}}$ order index of a given representation is defined as $l_{2n}(R) = \sum_{w\in \Lambda_{R}} (w,w)^{n}$
where $w$ are the weights in the weight lattice $\Lambda_{R}$ of the representation, and the inner product $(\cdot,\cdot)$ is the standard Killing metric on the weight space. The coefficient $a_{R}$ is the index $l_{2}(R)$, normalised such that it gives 1 on the fundamental representation. It is noteworthy that $l_4(R)=0$ for $A_1, A_2, G_2, F_4, E_6, E_7, E_8$.
For the discussion of trace relations and indices in the case of $D_4$, we refer to \cite{Okubo:1981td}. See also \cite{Becker:2026etl} for a recent review. In this case we define we need only the coefficients tabulated below.
\be
\centering
\begin{tabular}{ccccc}
\hline
dim & $a_R$ & $b_R$ & $c_R$ & $d_R$ \\
\hline
${\bf 1}$ & 0 & 0 & 0 & 0 \\
${\bf 8_v}$ & 1 & 1 & 0 & 0 \\
${\bf 8_c}$ & 1 & $-\frac12$ & $\frac38$ & -12 \\
${\bf 8_s}$ & 1 & $-\frac12$ & $\frac38$ & 12 \\
${\bf 28}$ & 6 & 0 & 3 & 0 \\
\hline
\end{tabular}
\label{D4indices}
\ee

For models in which Green-Schwarz mechanisms works to remove the local anomalies, the anomaly polynomial $I_8$ factorizes as 
\be
\frac{1}{2\pi} I_8 = \frac{1}{2} \eta_{\alpha\beta} Y^\alpha Y^\beta\ ,\qquad \alpha=+,-\ ,
\label{I8}
\ee
where $\eta_{\alpha\beta}$ is the invariant tensor of $SO(n_T,1)$, and for the gauge group \eq{g1} 
\be
Y^\alpha = \frac{1}{16\pi^2} \Big[\frac12  a^\alpha \tr R^2 +\sum_{i=1}^{n} b_i^\alpha \left(\frac{2}{\lambda_i} \tr F_{i}^2\right) + 2c^\alpha\, F_{R}^{2}\Big]  \ ,
\label{X4}
\ee
and the normalization factors $\lambda_i$ for $i=1,...,n$ are fixed such that the smallest topological charge of an embedded $SU(2)$ instanton is $1$, and they are listed below.
\medskip
\begin{align}
&& A_n && B_n && C_n && D_n && E_6 && E_7 && E_8 && F_4 && G_2 
\nn\\
\lambda && 1 && 2 && 1 && 2 && 6 && 12 && 60 && 6 && 2
\label{cn}
\end{align}
The 4-form $Y^{\alpha}$ can be written in terms of characteristic classes as follows.
\be
Y^{\alpha} = -\frac{1}{4}a^{\alpha}p_{1} + \sum_{i} b_{i}^{\alpha} c_{2}^{i} + \frac{1}{2}c^{\alpha} c_{1}^{2}\ ,
\label{YC}
\ee
where
\be
p_{1} = -\frac{1}{8\pi^{2}}\tr R^{2} \, \qquad c_{2}^{i} = \frac{1}{8\pi^{2}\lambda_{i}}\tr F_{i}^{2} \ , \qquad c_{1} = \frac{1}{2\pi}F_{R}\ .
\ee
Here $p_1$ is the first Pontryagin form of the tangent bundle, $c_1$ is the first Chern class form of the $U(1)_R$ bundle and $c_2^i$ is the second Chern form associated to the Lie algebra ${\mathfrak g}_i$. 

For details of the action, we refer to \cite{Nishino:1986Dc,Ferrara:1997gh,Riccioni:1998th,Bossard:2024ffp}. For later purposes, we record here the bosonic Lagrangian in the absence of the hypermultiplet sector, which takes the form \cite{Riccioni:2001bg,Bossard:2024ffp}
\begin{align}
e^{-1}\mathcal{L} = &\frac14 R - \frac14\partial_\mu \phi \partial^\mu \phi -\frac{1}{12} e^{2\phi} H_{\mu \nu \rho} H^{\mu \nu \rho} -\frac14 j_\alpha v^\alpha_i \Tr_i(F_{\mu\nu}F^{\mu\nu}) -\frac14 j_\alpha v^\alpha F_{R \mu\nu}F_R^{\mu\nu} 
\nonumber \\
&  + \frac18 e^{-1}\epsilon^{\mu \nu \rho \sigma \tau \lambda} B_{\mu \nu} \left( v^-_i  \Tr_i(F_{\rho \sigma} F_{\tau\lambda}) + v^- F_{R\rho \sigma} F_{R\tau\lambda} \right) 
\nn\w2
& - \frac12 e^{-1}\epsilon^{\mu \nu \rho \sigma \tau \lambda} v^+_i v^-_j X^i_{\mu\nu\rho} X^j_{\sigma\tau\lambda} - \frac{1}{j_\alpha v^\alpha}\ ,
\label{ga}
\end{align}
where $H=dB-v_i^+X_i -v^+X_R$ with Chern-Simons forms $X_i=\Tr_i(AdA+\frac23 A^3)$ and $X_R=A_RdA_R$, with $\Tr (XY) = \delta_{IJ} X^I Y^J$, and $v^+v^- =b^+ b^-/(32\pi^3)$ as can be deduced from the cancellation of the anomaly $I_6^1 + \delta {\cal L}_{GS}=0$ for all group factors, with the  descent equations $I_8=dI_7^0$ and $\delta I_7^0=dI_6^1$ understood.  Furthermore $j^\alpha$ is an  $SO(1,1)$ vector defined by
\begin{align}
 & j\cdot j=+1\ ,\quad e\cdot e=-1\ ,\quad e\cdot j=0\ ,
 \nn\\
& -e_\alpha e_\beta +j_\alpha j_\beta = \eta_{\alpha\beta} = \begin{pmatrix} 0 &1\\1&0\end{pmatrix}\ .
\label{bv2}
\end{align}
where we have introduced the orthogonal vector $e^\alpha$ so that $(e^\alpha,j^\alpha)$ forms a $SO(1,1)$ group element. These vectors can be chosen as 
\be
e^\alpha = \frac{1}{\sqrt2}  \begin{pmatrix} e^{-\phi} \\ -e^\phi \end{pmatrix}\ ,\qquad j^\alpha = \frac{1}{\sqrt2}  \begin{pmatrix} e^{-\phi} \\ e^\phi \end{pmatrix}\ .
\label{eje}
\ee
The metric on the tensor moduli space $\bbR^+$ is manifestly positive given by 
$ G_{\alpha\beta} = e_\alpha e_\beta +j_\alpha j_\beta$, while the ghost-free conditions for the vector fields take the form
\be
j_\alpha b^\alpha >0\ ,\qquad  j_\alpha c^\alpha >0\ .
\label{gf1}
\ee
While in this paper we are not considering higher derivative extension of 6D supergravity, the positivity of the Riemann-squared term that arises is ensured by 
\be
j_\alpha a^\alpha <0 \ .
\label{conj}
\ee
This condition is expected to hold but it remains to be proven \cite{Cheung:2016wjt,Hamada:2018dde,Caron-Huot:2021rmr,Bobev:2021qxx}. In this paper we shall impose both  \eq{gf1} by and \eq{conj}. It is worth noting that taking letting $j\to -j$ preserves the $SO(1,1)$ conditions as well as supersymmetry of the action. This amounts to letting $(a,b,c) \to -(a,b,c)$ which preserve the anomaly polynomial $I_8$. This may be used to deal with vector kinetic terms in which $b$ and/or $c$ have all negative components; see, for example, the $b$ vector for $E_7$ for model 1 below. However, it is also be noted that this process changes not only the sign of the vector kinetic terms but also $j\cdot a$ which arises in front  of the Riemann-squared (or Gauss-Bonnet) invariant) which is yet to be constructed for the most general gauged anomaly free models. 

In the rest of this section, we note the field content and anomaly polynomials of 9 known gauged supergravities, as well as 13 new ones we have found, with $n_{T} = 1$, $\Uni{1}_{R}$ symmetry gauging, and non-trivial Yang-Mills sector containing only semi-simple groups, where each simple factor has rank greater than 1.%
\footnote{There is only one model with only $U(1)_R$ symmetry which is anomaly free by Green-Schwarz mechanism \cite{Suzuki:2005vu}, but it has global anomaly since it does not satisfy the condition \eq{ncs}.}
For a long time only 3 such models, referred to as the ``old models" below,  were known. More recently, 6 new models, referred to as the ``recent models" were found in \cite{Becker:2023zyb}, where gauge groups of the form $G_1\times G_2 \times G_3\times U(1)_R$ with $G_i$ from the set $G_2, F_4, E_6, E_7, E_8, A_9,A_{10}, B_N, C_N, D_N\ , \ \mbox{for}\ 5\le N\le 10$ were considered. Remarkably the models found had no hyperino singlets. Here, we have expanded the search by considering gauge groups that are products of up to 4 group factors, and we include $A_N$ for $ 3\le N\le 8, B_3, B_4, C_3, C_4, D_4$ in the list of groups. However, we do not consider the most general groups formed from the expanded set, nor the most general representation content. Nevertheless, we have found that the 24 models we have arrived at arise among a huge number of possible gauge group and matter field content combinations, and this number is far smaller than the number of anomaly-free {\it ungauged} models.

\subsection*{\it The old models}

The first three models were found in \cite{Avramis:2005hc,Randjbar-Daemi:1985tdc,Avramis:2005qt} and have been studied in further detail in \cite{Randjbar-Daemi:2004bjl,Pang:2020rir}. Their gauge groups, representation content, and the anomaly coefficient two-vectors  $ \left(a^\alpha,b_{i}^\alpha,c^\alpha \right)$ are as follows
\footnote{Following \cite{Kumar:2010ru}, in reading off $(a,b,c)$ from $I_8$, the freedom in sending $Y\to -Y$ is used to set $a=(-2,-2)$.}
(For convenience, we now use the subscript $i$ to denote the rank of the associated simple gauge group factor.) 
\begin{align}
1. & \quad  \mathrm{E}_{6} \times \mathrm{E}_{7} \times \Uni{1}_{R}\ ,\quad  (\mathbf{1},\mathbf{912})_{0}\ ,
\nn\\
&\quad a = (-2,-2)\ , \quad b_{6} = (1,-3)\ , \quad b_{7} = (3,9) \ , \quad c= (2,-18)\ ,
\label{m1}
\w2
2. &\quad \mathrm{F}_{4}\times\Sp{9}\times \Uni{1}_{R}\ ,\quad \ (\mathbf{52},\mathbf{18})_{0}\ ,
\nn\\
& \quad a=(-2,-2) \ , \quad b_{4} = (2,10) \ , \quad b_{9} = (1,-\tfrac{1}{2}) \ , \quad c = (2,-19)\ .
\label{m3}
\end{align}
As noted in \cite{Randjbar-Daemi:1985tdc}, given that only the adjoint representation of $E_6$ is present, considering the global structure of the gauge group, it can be chosen to be the quotient $\mathrm{E}_{6}/\bbZ_{3}$. This group was considered in \cite{Pang:2020rir}, where its global anomalies as well as the global anomalies of the $F_4 \times Sp(9)\times U(1)_R$ were studied. We shall come back to this point in Section 4.

\subsection*{\it The recent models}

Recently, we found six new gauged supergravities in \cite{Becker:2023zyb}. Their consistency is yet to be studied and so we will also consider these models in the following sections. As far as the local anomaly freedom is concerned, in which case the global structure of the group does not play a role, the gauge group, representation content and the anomaly coefficients for these models are as follows.
\begin{align}
3. &\quad \mathrm{E}_{7}\times \mathrm{E}_{8}\times \mathrm{Spin}(20)\times \Uni{1}_{R}\ ,\quad \ (\mathbf{56},\mathbf{1},\mathbf{20})_{0} + (\mathbf{1},\mathbf{1},\mathbf{512})_{0}\ ,
\nn\\
& \quad a=(-2,-2) \ , \quad b_{7} = (1,6) \ , \quad b_{8} = (1,-6) \ , \quad b_{20} = (1,6) \ , \quad c=(2,-48)\ ,
\label{m4}
\w2
4. &\quad \mathrm{E}_{6}\times\mathrm{E}_{6}\times \Sp{5}\times \Uni{1}_{R}\ ,\quad \ (\textbf{78},\textbf{1},\textbf{10})_{0} + (\textbf{1},\textbf{1},\textbf{132})_{0}\ ,
\nn\\
& \quad a = (-2,-2)\ , \quad b_{6} = (2,6) \ , \quad b'_{6} = (1,-3) \ , \quad b_{5} = (1,3) \ , \quad c=(2,-18)\ ,
\label{m5}
\w2
5. &\quad  \mathrm{E}_{7} \times \SU{10} \times \SU{10} \times \Uni{1}_{R}\ ,\quad \ (\textbf{1},\textbf{10},\textbf{45})_{0} + (\textbf{1},\textbf{252},\textbf{1})_{0}\ ,
\nn\\
& \quad a=(-2,-2) \ ,\quad b_{7} = (1,-4) \ ,\quad b_{10} = (1,4) \ , \quad b'_{10} = (1,4) \ , \quad c= (2,-28)\ ,
\label{m6}
\w2
6. &\quad \mathrm{F}_{4}\times \mathrm{Spin}(13) \times \Sp{7} \times\Uni{1}_{R} \ ,\quad \ (\mathbf{26},\mathbf{1},\mathbf{14})_0+(\mathbf{1},\mathbf{13},\mathbf{14})_0+(\mathbf{1},\mathbf{1},\mathbf{350})_0 
\nn\w2
& \hspace{6.4cm} + (\mathbf{1},\mathbf{64},\mathbf{1})_0\ ,
\nn\\
& \quad a=(-2,-2) \ , \quad b_{4} = (1,1) \ , \quad b_{13} = (1,-1) \ , \quad b_{7} = (1,2) \ , \quad c = (2,-20)\ ,
\label{m7}
\w2
7. &\quad  \Sp{5} \times \Sp{6} \times \mathrm{Spin}(12) \times \Uni{1}_{R}\ ,\quad \ (\mathbf{10},\mathbf{1},\mathbf{12})_0+ (\mathbf{44},\mathbf{12},\mathbf{1})_0 +(\mathbf{1},\mathbf{1},\mathbf{32})_0 
\nn\w2
& \hspace{6.8cm} +(\mathbf{1},\mathbf{208},\mathbf{1})_0 \ ,
\nn\\
& \quad a=(-2,-2) \ ,\quad b_{5} = (1,\tfrac{5}{2}) \ , \quad b_{6} = (1,\tfrac{3}{2}) \ , \quad b_{12} = (1,-\tfrac{3}{2}) \ ,\quad c = (2,-17) \ ,
\label{m8}
\w2
8. &\quad \Sp{5} \times \Sp{6} \times \mathrm{Spin}(13) \times \Uni{1}_{R}\ ,\quad \ (\mathbf{10},\mathbf{1},\mathbf{13})_0 + (\mathbf{10},\mathbf{65},\mathbf{1})_0 + (\mathbf{132},\mathbf{1},\mathbf{1})_0 \ ,
\\
& \quad a=(-2,-2) \ , \quad b_{5} = (1,3) \ , \quad b_{6} = (1,2) \ , \quad b_{13} = (1,-2) \ , \quad c = (2,-18)\ .
\label{m9}
\end{align}
These coefficients were found using the conventions that the $\SO{1,n_{T}}$ invariant tensor $\eta$ takes the off-diagonal form given in \eq{eje}. 

\subsection*{\it New models}

In this paper, we are presenting 15  new R-symmetry gauged models in which $n_T=1$ and all local anomalies are canceled by the Green-Schwarz mechanism. The gauge groups, matter content and the anomaly coefficients of these models are as follows\footnote{We thank Qi You for pointing out a typo in an earlier version of model 18.}
\begin{align}
9. & \quad \mathrm{E}_7\times \mathrm{Sp(10)}\times \Uni{1}_{R}\ ,
\qquad  (\mathbf{1},\mathbf{1120})_0 + (\mathbf{56},\mathbf{1})_0\ ,
\nn\\
& \quad a=(-2,-2)\ ,\quad b_7= (1,-7/2)\ ,\quad b_{10}=(1,7/2)\ , \quad c=(2,-29)\ ,
\label{m10}
\w4
10. &\quad \mathrm{SU}(8)\times \mathrm{Spin}(8)\times \Uni{1}_{R}\ ,\quad 
({\bf 336}, {\bf 1})_0\ ,
\nn\\
&\quad a=(-2,-2)\ ,\quad b_8=(4,8)\ ,\quad b_8=(1,-2)\ ,\quad c=(2,-8)\ ,
\label{m11}
\w4
11. & \quad \mathrm{Spin(17)}\times \mathrm{Sp(9)}\times \Uni{1}_{R}\ ,
\qquad  (\mathbf{1},\mathbf{798})_0 + (\mathbf{17},\mathbf{18})_0\ ,
\nn\\
& \quad a=(-2,-2)\ ,\quad b_7= (1,-2)\ ,\quad b_{10}=(1,3)\ , \quad c=(2,-26)\ ,
\label{m12}
\w4
12. &\quad \mathrm{E}_{7}\times \mathrm{Spin}(15)\times \mathrm{Sp}(3)\times \Uni{1}_{R}\ ,\qquad   ({\bf 1}, {\bf 15},{\bf 14}^\prime)_0+   
    ({\bf 133}, {\bf 1},{\bf 6})_0\, ,
    \nn\\ 
    & \quad a=(-2,-2)\ , \quad b_7=(2,4)\ ,\quad b_{15}=(1,-2)\ ,\quad b_3=(1,7)\ ,\quad c=(2,-22) \ ,
    \label{m13}
\w4
13. &\quad \mathrm{F}_{4}\times \mathrm{Sp}(3)\times \mathrm{Sp}(6)\times \Uni{1}_{R}\ ,
\qquad ({\bf 1}, {\bf 14},{\bf 12})_0+  ({\bf 26}, {\bf 1},{\bf 12})_0+   
    ({\bf 52}, {\bf 6},{\bf 1})_0\ ,
\nn\\
& \quad a=(-2,-2) \ , \quad  b_4 = (2,4) \ , \quad b_3 = (1,5/2) \ , \quad b_{6} = (1,-1/2) \ , \quad c = (2,-13)\ ,
\label{m14}
\w4
14. &\quad \mathrm{G}_{2}\times \mathrm{SU}(6)\times \mathrm{Spin}(12)\times \Uni{1}_{R}\ ,
\qquad ({\bf 7}, {\bf 1},{\bf 32})_0  +  ({\bf 1}, {\bf 1},{\bf 352})_0  +  ({\bf 1}, {\bf 6}, {\bf 12} ) \ ,
\nn\\
& \quad a=(-2,-2) \ , \quad  b_2 = (1,1) \ , \quad b_6 = (1,-1) \ , \quad b_{12} = (3,5) \ , \quad c = (2,-10)\ ,
\label{m15}
\w4
15. &\quad \mathrm{SU}(8)\times \mathrm{Sp}(5)\times \mathrm{Spin}(10)\times \Uni{1}_{R}\ ,\quad 
({\bf 1}, {\bf 10},{\bf 45})_0+ ({\bf 1}, {\bf 110},{\bf 1})_0+  ({\bf 8}, {\bf 1},{\bf 16})_0\ ,
\nn\\
&\quad a=(-2,-2)\ ,\quad b_8=(1,-1)\ ,\quad b_5=(1,1)\,\quad b_{10}=(2,6)\ ,\quad c=(2,-14)\ ,
\label{m16}
\w4
16. &\quad \mathrm{SU}(6)\times \mathrm{Spin}(8)\times \mathrm{Spin}(8)\times \Uni{1}_{R}\ ,\quad 
({\bf 20}, {\bf 28},{\bf 1})_0 + ({\bf 56}, {\bf 1},{\bf 1})_0 \ ,
\nn\\
&\quad a=(-2,-2)\ ,\quad b_6=(3,6)\ ,\quad b_8^{(1)}=(3,6)\,\quad b_{8}^{(2)}=(1,-2)\ ,\quad c=(2,-8)\ ,
\label{m17}
\w4
17. &\quad \mathrm{Sp}(4)\times \mathrm{Sp}(4)\times \mathrm{Sp}(5)\times \Uni{1}_{R}\ ,\quad \ (\mathbf{1},\mathbf{48},\mathbf{1})_{0} + (\mathbf{36},\mathbf{1},\mathbf{10})_{0}+ (\mathbf{42},\mathbf{8},\mathbf{1})_{0}\ ,
\nn\\
& \quad a=(-2,-2) \ , \quad b_4^{(1)} = (2,6) \ , \quad b_4^{(2)} = (1,1/2) \ , \quad b_{5} = (1,-1/2) \ , \quad c = (2,-11)\ ,
\label{m19}
\\
18. & \quad  \mathrm{E}_7\times \mathrm{E}_7\times \mathrm{E}_7 \times \mathrm{Spin}(8)\times \Uni{1}_{R}\ ,
\nn\\
&\quad  (\mathbf{1},\mathbf{1},\mathbf{56},\mathbf{8}_s)_0 + (\mathbf{1},\mathbf{56},\mathbf{1},\mathbf{8}_c)_0 +(\mathbf{56},\mathbf{1},\mathbf{1},\mathbf{8}_v)_0\ ,
\nn\\
& \quad a=(-2,-2)\ , \quad b_7^{(1)}=b_7^{(2)}=b_7^{(3)}=(1,0)\ ,\quad  b_{8}=(1,12)\ , \quad c=(2,-36)\ ,
\label{m20}
\w4
19. & \quad  \mathrm{F}_4\times \mathrm{F}_4\times \mathrm{F}_4 \times \mathrm{Sp}(5)\times \Uni{1}_{R}\ ,
\nn\\
&\quad  (\mathbf{1},\mathbf{1},\mathbf{1},\mathbf{132})_0 + (\mathbf{1},\mathbf{1},\mathbf{26},\mathbf{10})_0 +(\mathbf{1},\mathbf{26},\mathbf{1},\mathbf{10})_0
+ (\mathbf{26},\mathbf{1},\mathbf{1},\mathbf{10})_0\ ,
\nn\\
& \quad a=(-2,-2)\ , \quad b_4^{(1)}=b_4^{(2)}=b_4^{(3)}=(1,0)\ ,\quad  b_{10}=(1,3)\ , \quad c=(2,-18)\ ,
\label{m21}
\w4
20. & \quad \mathrm{E}_7\times \mathrm{Spin}(12)\times \mathrm{Spin}(12)\times \mathrm{Spin}(12) \times \Uni{1}_{R}\ ,
\nn\\
& \quad  (\mathbf{1},\mathbf{1},\mathbf{32},\mathbf{12})_0 + (\mathbf{1},\mathbf{12},\mathbf{1},\mathbf{32})_0 +(\mathbf{1},\mathbf{32},\mathbf{12},\mathbf{1})_0\ ,
\nn\\
& \quad a=(-2,-2)\ ,\quad b_7= (1,-4)\ ,\quad b_{12}^{(1)}=b_{12}^{(2)}=b_{12}^{(3)}=(1,4)\ , \quad c=(2,-28)\ ,
\label{m22}
\w4
21. &\quad \mathrm{F}_{4}\times \mathrm{Spin}(9)\times \mathrm{Sp}(3)\times \mathrm{Sp}(6)\times \Uni{1}_{R}\ ,
\nn \\
& \quad  ({\bf 1}, {\bf 1},{\bf 6},{\bf 65})_0+  
      ({\bf 1}, {\bf 9},{\bf 14}^\prime,{\bf 1 })_0+  
      ({\bf 1}, {\bf 16},{\bf 1},{\bf 12})_0+  
      ({\bf 26}, {\bf 1},{\bf 6},{\bf 1 })_0 \ ,
      \nn \\
      & a=(-2,-2)\ ,\quad b_4=(1,-1)\ ,\quad b_9=(1,1)\ ,\quad 
      b_3 =(1,4)\ ,\quad b_6=(1,1)\ ,\quad c=(2,-16) 
\label{m23}
\w4
22. &\quad \mathrm{Spin}(9)\times \mathrm{Spin}(9)\times \mathrm{Sp}(4)\times \mathrm{Sp}(5) \times \Uni{1}_{R}\ ,
\nn\\
& 
\quad  ({\bf 1 }, {\bf 1},{\bf 1 }, {\bf 110})_0+  
 ({\bf 1}, {\bf 1},{\bf  27}, {\bf 10})_0+  
 ({\bf 1 }, {\bf 9},{\bf 1}, {\bf 10})_0+ 
({\bf 1}, {\bf 16 },{\bf 8 }, {\bf 1})_0+  
  ({\bf 9}, {\bf 1},{\bf 1}, {\bf 10})_0+  ({\bf 16}, {\bf 1},{\bf 8 }, {\bf 1})_0\ ,
\nn\\
& \quad a=(-2,-2)\ ,\quad b_9^{(1)}=b_9^{(2)}=(1,0)\ ,\quad b_4=(1,2)\ ,\quad b_5= (1,1)\quad c=(2,-14)\ .
\label{m24}
\end{align}
In computing the anomaly polynomials, it is useful to bear in mind that the ${\bf 56}$ and ${\bf 912}$ of $E_7$, the ${\bf 2N}$ of ${\rm Sp}(N)$ and $2^{\frac{[N+1}{2}]-1}$ dimensional representations of ${\rm SO}(N)$ for $N=3,4,5\ {\rm mod}\ 8$ are pseudoreal. The models 9 and 11 were missed in \cite{Becker:2023zyb}, while all the other new models either involve certain representations or four group factors in addition to $U(1)_R$, which were not considered in \cite{Becker:2023zyb}. 

\subsection*{\it Connectivity of the groups}

All of the groups above are simply connected. For some of the models with the representation content given above, it is possible to consider quotient groups by modding out appropriate elements of the center groups. For example, in model 1 the group $E_6$, and in model 5 one or both of the $E_6$ groups can be modded out by $\bbZ_3$. Other examples are as follows: In model 4, the group $E_7\times \mathrm{Spin}(20)$, and in model 8 the group $\mathrm{Spin}(12)\times \Sp{5}$ can be modded out by a diagonal action of the centers $\bbZ_2$, and in model 6, the group $\SU{10} \times \SU{10}$ can be modded out by a suitably identified diagonal action of $\bbZ_5$.\footnote{We thank Guillaume Bossard for helpful discussions  in determining the appropriate quotient groups.} Such quotient groups will not be simply connected, and we shall not consider these models.

\section{Charge lattices and global anomalies}
\label{sec:global_anomalies}

Beyond local anomaly freedom, there are other consistency requirements on the anomaly polynomial associated to global anomalies. These results were summarised in \cite{Monnier:2017oqd}, and they imply that the coefficients in the factorised anomaly polynomial must define a suitable integral bilinear form on the charge lattice of dyonic strings. The arguments leading to the global anomaly freedom conditions begin with the observation that the factor
\be
\exp \left(-2\pi i \cdot \frac12 \int \eta_{\alpha\beta} B^\alpha Y^\beta \right)\ ,
\ee
is included in the ``path integral" of the theory. Next, the {\it completeness hypothesis} is noted, for which supporting arguments have been given in \cite{Polchinski:2003bq,Banks:2010zn,Hellerman:2010fv}. It conjectures that all the charges allowed by the Dirac quantization condition are present in consistent quantum gravity theories. In \cite{Monnier:2017oqd}, a stronger version of this hypothesis is assumed.  It additionally states that a consistent theory of supergravity may be put on an arbitrary spin manifold, and that any smooth gauge field configuration should be allowed in the 
supergravity ``path integral". Next, it is observed, that on a general spacetime endowed with a general gauge bundle, $Y$ is nontrivial and carries background charges, which have to be canceled by introducing background strings, whose existence is ensured by the completeness hypothesis. Thus, the generalized completeness conjecture allows the authors of \cite{Monnier:2017oqd} to choose freely the spacetime and the gauge bundle when examining the consistency of six-dimensional theories. Indeed, Euclideanized spacetimes of the form $S^2 \times S^4$ and $CP^3$ have been considered in \cite{Monnier:2017oqd} in deriving some new anomaly freedom constraints, as will be recalled below. 

The cancellation of the charge implied by the presence of $Y$ is possible when the integral of $Y$ along any integral 4-cycle $\Sigma_4$ yields an element of the string charge lattice $\Lambda_S$. This statement explicitly reads
\be
\int_{\Sigma_4} Y = a \int_{\Sigma_4} \frac14 p_1 - \sum_{i} b_i \int_{\Sigma_4} c_2^i + \frac12 c^\alpha \int_{\Sigma_4} c_1^2 \in \Lambda_S\ ,
\label{sqc}
\ee
which is referred to as the ``string quantization condition".  This condition was checked for models \eq{m2}, \eq{m3} and \eq{m1} in \cite{Pang:2020rir} where it was found that only the model \eq{m3} among them was consistent with the additional criteria. We will briefly review these criteria here and apply them to the model \eq{m1} and models \eq{m4}-\eq{m9} as well, and show that they are satisfied. As a by product, we will find a powerful constraint \eq{ncs} on the number of vector multiplets in any R-symmetry gauged models with $n_{T}=1$. 

\subsection{Consistency conditions on anomaly coefficients from the charge lattice}\label{sec:charge_lattice_consistency}

The conditions for consistency of an $N=(1,0)$ supergravity theory in 6-dimensions with $n_{T}$ tensor multiplets and \mbox{\it simply connected} gauge group $G$, as summarised in \cite{Monnier:2017oqd}, are as follows.
\begin{align}
(i) &\quad  \frac{1}{2\pi i} I_{8} = \tfrac{1}{2}\eta_{\alpha\beta}Y^{\alpha}Y^{\beta}
\label{c1}
\w2
(ii) &\quad  \mbox{The charge lattice for dyonic strings} \ \Lambda_{S} \ \mbox{is unimodular.} 
\label{c2}
\w2
(iii) & \quad a\in \Lambda_{S} \ \mbox{is a characteristic element.} 
\label{c3}
\w2
(iv) & \quad b_{i}^{\alpha}, \, \tfrac{1}{2}c^{\alpha} \in \Lambda_{S}\ .
\label{c4}
\w2
(v) & \quad  \Omega^{\text{spin}}_{7}(BG) = 0\ .
\label{cg}
\end{align}
Condition (i) is the statement that the local anomalies can be cancelled through a Green-Schwarz-Sagnotti mechanism as in Sec.~\ref{sec:gauged_review}. 

Condition (ii) states that the lattice comes equipped with an inner product ($x\cdot y = x^{\alpha}\eta_{\alpha\beta}y^{\beta}$ for $x,\, y\in \Lambda_{S}$), and the lattice is self dual. Alternatively, given a basis $e_{a}$ of $\Lambda_{S}$, the Gram matrix $G_{ab} = e_{a}\cdot e_{b}$ has determinant $-1$.\footnote{In general, a unimodular lattice can have a Gram matrix with determinant $\pm1$. However, in 6-dimensional supergravity the charge lattice has the same signature as the tensor multiplet moduli space which is $(1,n_{T})$. Hence, for $n_{T}=1$ we require $\det G_{ab} = -1$.} This condition comes from requiring that the wave function of a dyonic string of charge $Q_{1}$ is well-defined when rotated $2\pi$ around another dyonic string of charge $Q_{2}$ \cite{Deser:1997se,Seiberg:2011dr}.\footnote{One must be careful to ensure that the motion of the two strings forms a contractible loop in phase space. Hence a suitable `double-pass' is required. See \cite{Deser:1997se} for more details.} When this happens, the wave function picks up a phase, in suitable units, of $2\pi Q_{1}\cdot Q_{2}$. Hence, to be well-defined, we need $Q_{1}\cdot Q_{2} \in \bbZ$, which is the statement that $\Lambda_{S}$ is unimodular. 
The unimodularity of the charge lattice can be conveniently tested as follows. Given any elements $x,y\in \Lambda_{S}$ of a unimodular lattice, we consider the matrix
\be
    \mathcal{M}(x,y) = 
    \left( \begin{array}{cc}
        x\cdot x & x\cdot y \\
        y\cdot x & y\cdot y
    \end{array} \right)\ ,
\ee
where $x^\alpha= n_a e^\alpha_a$ and $y^\alpha= m_a e^\alpha_a$ and $n_a, m_a$ are integers. Taking the determinant of this matrix gives
\begin{equation}\label{eq:det_-n2}
\det \mathcal{M}(x,y) = (\det G) \, (\epsilon^{ab} n_a m_b)^2 = -p^2 \quad \mbox{for some}\quad p\in \mathbb{Z}\ .
\end{equation}
This implies that if $a,b_{i},\tfrac{1}{2}c$ are elements of a unimodular charge lattice, then $-\det \mathcal{M}$, formed by the choice of any pair of vectors, must be a square integer. This condition goes beyond the fact that $a\cdot a, a\cdot b_i, b_i\cdot b_j \in \mathbb{Z}$ for all $i, j$, as was shown in \cite{Kumar:2010ru} to follow remarkably from the factorization of the anomaly polynomial $I_8$ alone. 

Condition (iii) simply states that
\begin{equation}
a\cdot x = x\cdot x \mod 2 \quad \forall\, x\in \Lambda_{S}\ .
\label{cec}
\end{equation}

This argument is analogous to the Dirac quantisation of electromagnetic charges in 4 dimensions.  
Condition (iv) can be reformulated in terms of cohomology groups over the classifying space $BG$ of the simply connected gauge group $G$. See \cite{Monnier:2017oqd,Pang:2020rir} for more details. 
Conditions (iii) and (iv) were derived using the generalised completeness principal and the consideration of the string charge quantisation condition \eq{sqc} for different spin manifolds \cite{Monnier:2017oqd}.
Condition (v) says that the spin cobordism group of the classifying space $BG$ of the gauge group $G$ in degree 7 must be trivial. This condition follows from the requirement  that the Green-Schwarz anomaly canceling term is globally well defined on generic backgrounds \cite{Monnier:2018nfs}. 

Prior to the consideration of the spin cobordism group in relation to the global anomaly freedom, the role of the sixth homotopy group $\pi_6(G)$ was considered in theories that admit the Green-Schwarz anomaly cancellation mechanism \cite{Tosa:1989qm} and conditions were derived on the anomaly coefficients $b^\alpha$ for the absence of global anomaliesir cancellation \cite{Bershadsky:1997sb}. The only such cases have $\pi_6(SU(2))={\mathbb Z}_{12}, \pi_6(SU(3))={\mathbb Z}_6$ and $\pi_6(G_2)={\mathbb Z}_3$. It is important to note that even for these groups, as long as the conditions \eq{c1}--\eq{cg} are satisfied, the ones that arise from the nonvanishing of $\pi_6(G)$ automatically follow. This point has been demonstrated in great detail in \cite{Lee:2020ewl,Davighi:2020kok}.

\subsection{Conditions on \texorpdfstring{$n_V$}{nV} from unimodularity of the charge lattice}\label{sec:nV_constraints}

In the case that $n_{T} = 1$, we can use the requirement that $a$ and $\tfrac{1}{2}c$ must be elements of a unimodular lattice to put strict constraints on the dimension of the gauge group $n_{V}$. In particular, we can use \eqref{eq:det_-n2} to show that, for $\Uni{1}_{R}$ and $\Sp{1}_{R}$ gauging, we must have
\begin{align}
\Uni{1}_R: &  \qquad n_V\equiv 8\mod 12\ ,
\nn\w2
\Sp{1}_R: & \qquad \n_V =12 \ \mbox{or}\ 96\ . 
\label{nvc}
\end{align}

Recalling that in our conventions, $n_{V}$ denotes the dimension of the entire gauge group, including the R-symmetry sector, one can check that all the models listed in Sec.~2 obey this condition. Next, we we proceed to prove the conditions \eq{nvc}. 
Working in a frame in which the $a$ and $c$ coefficients, and the inner product $\eta$, take the following form
\begin{equation}\label{ai}
a = (-2,-2) \ , \quad c = (c_{1},c_{2}) \ , \quad \eta = \left( \begin{array}{cc}
0  & 1 \\
1 & 0
\end{array} \right) \ ,
\end{equation}
we consider the matrix $\mathcal{M}(a, \tfrac{1}{2}c)$, given by
\begin{equation}
\mathcal{M} = \left( \begin{array}{cc}
a\cdot a & a\cdot (\tfrac{1}{2}c) 
\\
a\cdot (\tfrac{1}{2}c) & (\tfrac{1}{2}c)\cdot (\tfrac{1}{2}c)
\end{array} \right) = \left( \begin{array}{cc}
8 & -c_{1} - c_{2} 
\\
-c_{1} - c_{2} & \tfrac{1}{2}c_{1}c_{2}
\end{array} \right)
\end{equation}
This has determinant
\begin{equation}
\det \mathcal{M} = 4c_{1}c_{2} - (c_{1} + c_{2})^{2} = -(c_{1} - c_{2})^{2}\ .
\end{equation}
From \eqref{eq:det_-n2}, we know that this must be a negative square integer, and hence
\begin{equation}\label{aii}
    c_{1} - c_{2} \in  \bbZ \quad \Rightarrow \quad c_{2} = c_{1} + n \ , \ n \in \bbZ
\end{equation}
We now consider the $\Uni{1}_{R}$ and $\Sp{1}_{R}$ gauged cases separately.

\subsection{The case of $U(1)_R$}

From the factorisation of the anomaly polynomial, we find that
\begin{align}
    a\cdot c &= \tfrac{1}{6}(n_{V} - 20) \, , \\
    c \cdot c &= -\tfrac{1}{3}(n_{V} + 4) \, ,
\end{align}
while from (\ref{ai}) and (\ref{aii}) which are required by global anomaly considerations, we find
\begin{align}
    a\cdot c &= -2c_{1} - 2c_{2} = -4c_{1} - 2n\, ,\\
    c\cdot c &= 2 c_{1}(c_{1} + n) \, .
\end{align}
Solving for $c_1$ and using it in the $c\cdot c$ equation leads to a quadratic Diophantine equation whose solution is in the integers $n$ and $n_V$. The solutions are labeled by a non-negative integer $k\in \bbZ^\star$ as follows  
\be
c=(2,-1-k)\ , \quad\mbox{or}\quad c=(-1-k,2)\ ,\qquad n_V=8 +12 k\ .
\ee
This result can be used to rule out a great number of gauged models found in the literature with $\Uni{1}$ and $\SU{2}$ factors in the gauge group \cite{Avramis:2005hc,Suzuki:2005vu}. 

\subsection{The case of $Sp(1)_R$}

While the content of this paper focuses mostly on $\Uni{1}_{R}$ R-symmetry gauged models, we can perform similar analysis to the previous section for the case of $\Sp{1}_{R}$ gauging. To this end, it is sufficient to consider the gravitational and $Sp(1)_R$ sector of the anomaly polynomial which takes the form \cite{Suzuki:2005vu}
\begin{align}
I_8(R, F_R) &=  \frac{2\pi}{(16\pi^2)^2}\Big[ (\tr R^2)^2 +\frac{1}{12}  \left(n_V-12 \right) \tr R^2 \tr_{\bf 2}F_{R}^2
-\frac{1}{6}  (n_V+84) \left(\tr_{\bf 2}F_{R}^2\right)^2 \Big]
\nn\w2
&= \frac{2\pi}{(16\pi^2)^2} \times \frac12 \eta_{\alpha\beta} \left(\frac12 a^\alpha \tr R^2 + 2b^\alpha_R \tr_{\bf 2} F_R^2 \right)\left(\frac12 a^\beta \tr R^2 + 2b^\beta_R \tr_{\bf 2} F_R^2 \right)\ ,
\label{me1}
\end{align}
where $\tr_{\bf 2}$ is the trace over the $Sp(1)_R$ doublet representation. It follows that
\begin{align}
a\cdot a &= 8\ ,
\label{c1}\w2
a\cdot b_R &= \frac{1}{12}(n_{V} - 12) := p \ ,
\label{c2}\w2
b_R\cdot b_R &= -\frac{1}{12}(n_{V} + 84)=-(p+8)\ .
\label{c3}
\end{align}
Thus, the unimodularity condition becomes
\be
\det \left( \begin{array}{cc}
a\cdot a & a\cdot b_R \\
a\cdot b_R & b_R\cdot b_R
\end{array} \right) = \det \left( \begin{array}{cc}
8 & p \\
p & -(p+8)
\end{array} \right)=  -m^2\ ,\qquad m\in \mathbb{Z}\ ,\quad p\in \mathbb{Z}\ ,
\ee
which gives $(m-p-4)(m+p+4)=48$. Defining $A:=m-p-4$ and $B:= m+p+4$, we have $A+B=2m$ and $AB=48$. Thus we can write $A:=2r$ and $B:=2s$ and therefore $(r,s)=(1,12), (2,6), (3,4)$ or negatives of these pairs. This gives $s-r=11,4,1$ or their negatives. We also have $B-A=2(p+4)$ that gives $s-r=p+4= 11,4,1$ or their negatives. Since we also have $n_V=12(p+1)>0$, it follows that $p=7,0$. Recalling \eq{c2}, this gives  $n_V=12$ or $n_V=96$.

\section{Application of the global anomaly freedom criteria}

We would like to apply these conditions to the models summarised in Sec.~2, and hence we will restrict our analysis to the case $n_{T} = 1$ from now on. In each case we shall also require that the conditions \eq{gf1} hold in order to have ghost-free gauge field kinetic terms. We shall list the dilaton domains for 
these condition to be satisfied at the end of this section. While we will not require the conjectural condition \eq{conj}, we shall nonetheless comment on the models in which it is satisfied as well.

All the models under consideration have no local anomaly and so by definition, they satisfy condition (i) above. For conditions (ii)-(iv) it is enough, for the simply connected gauge groups, to show that $a,b_{i}, \tfrac{1}{2}c$ are elements of a unimodular charge lattice $\Lambda_{S}$, and that $a\in \Lambda_S$ is a characteristic element. As noted earlier, this implies that $-\det \mathcal{M}$, where $\mathcal{M}$ is formed by the choice of any pair of vectors $a,b_{i}, \tfrac{1}{2}c$, must be a square integer. 

The models 1 and 3 were shown to satisfy conditions (i)-(iv) in \cite{Pang:2020rir}\footnote{In \cite{Pang:2020rir} the version of model 1 with non-simply connected gauge group $(E_6/\bbZ_3) \times E_7 \times U(1)_R$ was shown to fail to satisfy a stronger condition given in \cite{Monnier:2017oqd}.}, and model 2 is not. In the latter case,
\begin{equation}
 {\rm Model}\ 2:\qquad    -\det \mathcal{M}(a,\tfrac{1}{2}c) = (\tfrac{44}{3})^{2} \ .
\end{equation}
This is not an integer squared so we conclude that the lattice defined by $a,b_{i},\tfrac{1}{2}c$ cannot be embedded in a unimodular charge lattice $\Lambda_{S}$.  Thus model 2 does not satisfy the global anomaly freedom criteria \cite{Pang:2020rir}. For the remaining models, namely models 4-24 , we find that $-\det\mathcal{M}$ are square integers, and therefore the charge lattices for them are unimodular. Note also that, for models $1,4,...,7,9,11-13,15-18,20-24$ the vectors $(a,b,c)$ have integer coefficient with respect to the basis, $ e_{1} = (1,0)$ and $e_2=(0,1)$ which implies an even integer lattice. 
In the case of models $3,8,10,14,19$, given that there are half-integer values arising in the b-coefficient, one can choose the basis \cite{Pang:2020rir} $e_1=(2,0)$ and $e_2=(0,1/2)$, in which case a point $x\in \Lambda_S$ can be written as $x=(2m+n, n/2)$ where $m,n\in \bbZ$, and this is an {\it odd charge lattice}. 

Having established that all models except model 2 have unimodular charge lattice, we next check the  condition (iii). This condition is trivially satisfied for models that have even charge lattices. In the case of models which have odd charge lattices, for any $x\in \Lambda_S$ as specified for odd charge lattice, we get $x\cdot a - x\cdot x=  -2m(n+2)-n(n+3) =0 \mod 2$, and therefore the condition (iii) in the form \eq{cec} is satisfied by these models as well. 

Next, we turn to the condition (v), which is satisfied for all the groups involved in the gauged models considered here, thanks to the result derived in the appendix by Tachikawa, which states that $\Omega^{\text{spin}}_{7}(BG)=0$ for 
\begin{equation}
G=\sG_1\times \sG_2\times \cdots \times \sG_n\ ,
\label{foo2}
\end{equation}
where $\sG_i$ is $U(1)$ or any simple simply-connected non-Abelian compact Lie group.  
Thus, we conclude that the gauged models 4-16 satisfy all the local and global  anomaly freedom criteria as formulated in \cite{Monnier:2018nfs}. 

There remains the conditions \eq{gf1} and \eq{conj}. It is easy to check that all models satisfy these conditions for an appropriate choice of a domain for the dilaton. 
Thus, we conclude that all models listed in section 3, except model 2,  pass all the tests considered in sections 2 and 3.  Finally, we have checked that the $Sp(1)_R$ gauge model of \cite{Avramis:2005hc} specified in \eq{ak1} satisfies the unimodular charge lattice condition as well, with unitary vector field kinetic term for $e^{\phi}>3$.

\section{Conclusions}

In this paper, we have presented two new R-symmetry gauged $6D, N=(1,0)$ supergravity models for $n_T=1$, which belong to a rare class, of which only 23 are now known at present that  are free from all local anomalies. We have analysed the consistency of all these models by employing the criteria as formulated in \cite{Monnier:2018nfs} in a unified manner, involving the considerations of not only the global gauge transformations but also the Dirac quantization conditions on dyonic string charges which are related to the anomaly coefficients appearing in the factored anomaly polynomials.  We have found that the models \eq{m1} and \eq{m3}-\eq{m24} pass all the global anomaly freedom tests, and these models have ghost-free Maxwell and Yang-Mills kinetic terms for a given dilaton domain, easily determined from the anomaly coefficients.
As a byproduct of our analysis, we proved that for any R-symmetry gauged theory with $n_{T}=1$, in which the anomaly polynomial factorizes thereby admitting Green-Schwarz mechanism for anomaly cancellation, the additional requirement that the anomaly coefficients lie in a unimodular string charge lattice requires that the number of vector multiplets, or equivalently the total dimension of the gauge group, must satisfy $n_{V} \equiv 8\mod 12$ for $\Uni{1}_{R}$ gauging, and remarkably $n_{V}=12$ or $n_V=96$ for $\Sp{1}_{R}$ gauging.

An important next step is the analysis of anomaly inflows in these models. Their high energy unitarity and causality properties also deserve investigation. Whether any one of these models can be embedded in string/M theory remains to be a challenging open problem. This may turn out to be impossible, or it may require yet to be discovered novel mechanisms. It will also be useful to understand at a deeper level the rarity of the R-symmetry gauged models with $n_T=1$ and nonabelian gauge groups, compared to the abundance of anomaly free models that are not R-symmetry gauged.

As noted in the introduction, in the R-symmetry gauged models the potential gives a cosmological constant upon setting all scalar fields equal to zero. The sign of this constant depends on the anomaly coefficients $(v, \tv)$ associated with the R-symmetry group.  It is noteworthy that if $\tv=0$, a maximally symmetric vacuum does not exist, while  6D de Sitter and  (A)dS$_4 \times S^2$ solutions arise for  $v,\tv >0$ \cite{Guo:2025mlb}. These solutions are non-supersymmetric. A half-supersymmetric Mink$_4\times S^2$ solution exists with or without $\tv$ turned on. The Kaluza-Klein spectra for the 4D de Sitter and 4D Minkowski cases have been recently been determined in \cite{Guo:2025mlb}, where the need for R-symmetry gauging in which the R-symmetry group is a diagonal subgroup of $U(1)\in Sp(n_H)$ and $U(1)_R$. Such ``diagonally gauged" models were found in \cite{Suzuki:2005vu}, and it would be interesting to investigate further their properties, including their global anomalies.

In this paper, we have restricted our attention to models with simply connected gauge groups, and gauged models coupled to a single tensor multiplet, i.e. $n_T=1$. Furthermore, we have considered gauge groups of the form $G_1\times \cdots \times G_n\times U(1)_R$ with certain restrictions on the rank of $G_i$ and the representations carried by the hyperfermions to make the search feasible, given the enormous number of possible gauge groups and matter representation contents. A future direction is to relax some of these assumptions. For example, a systematic search for local and global anomaly free gauged models for $n_T>1$ would be useful. Given that the increased number of tensor multiplets make it easier to cancel the local anomalies by Green-Schwarz-Sagnotti mechanism, it would be interesting if a bound can be put on $n_T$ for R-symmetry gauged models that are free from local and global anomalies.

\subsubsection*{Acknowledgments}

We thank Guillaume Bossard, Inaki Garcia-Etxebarria,  Hong Lu, Miguel Montero, Greg Moore, Yi Pang, 
Cumrun Vafa, Daniel Waldram and Qi You for helpful discussions. 
This  work is supported in part by the NSF grants PHY-2112859 and PHY-2413006.
The appendix was provided by Yuji Tachikawa, whose work is supported in part
by WPI Initiative, MEXT, Japan at Kavli IPMU, the University of Tokyo
and by JSPS KAKENHI Grant-in-Aid (Kiban-C), No.24K06883

\appendix

\begin{appendix}

\section{Computations of bordism groups}

\def\sG{\mathsf{G}}
In this appendix
%
we show that $\Omega^\text{spin}_7(BG)=0$ for
\begin{equation}
G=\sG_1\times \sG_2\times \cdots \times \sG_n\ ,
\label{foo}
\end{equation}
where $\sG_i$ is $U(1)$ or any simple simply-connected non-Abelian compact Lie group.
We note that the vanishing of this bordism group when there is only a single factor in \eqref{foo} 
was shown in various previous papers
using Atiyah-Hirzebruch or Adams spectral sequences,
see e.g.~\cite{Beaudry:2018ifm,Monnier:2018nfs,Davighi:2020kok,Lee:2020ewl,Lee:2022spd}.
The computation presented here is a simple generalization.
Below, the notation $\sG$ is reserved for a single factor,
whereas $G$ is used for products.

\subsection{$H^{d\le 8}(B\sG;\bZ)$ and $H^{d\le 7}(B\sG;\bZ_2)$  for a single factor $\sG$ }

\begin{table}
\[
\begin{array}{c|cccccccccccc}
    \sG \setminus d & 0 & 1 & 2 & 3 & 4 & 5 & 6 & 7 & 8 \\
    \hline
    \hline
    U(1)&\bZ & 0 & \bZ & 0 & \bZ & 0 & \bZ & 0 & \bZ\\
    \hline
    SU(3) & \bZ & 0 & 0 & 0 & \bZ & 0 & \bZ & 0 & \bZ\\
    SU(n\ge 4) & \bZ & 0 & 0 & 0 & \bZ & 0 & \bZ & 0 & \bZ^2\\
    \hline
    Sp(1) &  \bZ & 0 & 0 & 0 & \bZ & 0 & 0 & 0 & \bZ & \\
    Sp(n\ge 2) &  \bZ & 0 & 0 & 0 & \bZ & 0 & 0 & 0 & \bZ^2 & \\
    \hline
    Spin(7) &  \bZ & 0 & 0 & 0 & \bZ & 0 & 0 & \bZ_2 & \bZ^2 & \\
    Spin(8) &  \bZ & 0 & 0 & 0 & \bZ & 0 & 0 & \bZ_2 & \bZ^3 & \\
    Spin(n\ge 9) &  \bZ & 0 & 0 & 0 & \bZ & 0 & 0 & \bZ_2 & \bZ^2 & \\
    \hline
    G_2 &  \bZ & 0 & 0 & 0 & \bZ & 0 & 0 & \bZ_2 & \bZ & \\
     F_4,E_{6,7,8} &  \bZ & 0 & 0 & 0 & \bZ & 0 & 0 & \bZ_2 & \bZ  \\
\end{array}
\]
\caption{Integral cohomology groups $H^d(B\sG;\bZ)$ of classifying spaces $B\sG$ for $\sG$ either $U(1)$ or simple simply-connected compact Lie groups.\label{tab:integ-coho}}
\end{table} 

We first need $H^d(B\sG;\bZ)$ for $d\le 8$ as a starting point of the computation.
They are listed in Table~\ref{tab:integ-coho}.
Unfortunately there are no convenient mathematical references where these information are tabulated.
Here we quickly summarize how the entries of the table above can be filled in.
\begin{itemize}
\item
The results for $U(1)$, $SU(n)$ and $Sp(n)$ are standard.
In these cases, the entire integral cohomology is a free polynomial algebra
and is given by  \begin{equation}
\begin{aligned}
H^*(BU(1);\bZ)&\simeq\bZ[c_1]\, ,\\
H^*(BSU(n);\bZ)&\simeq\bZ[c_2,\ldots,c_n]\, ,\\
H^*(BSp(n);\bZ)&\simeq\bZ[q_1,\ldots,q_n]\, .
\end{aligned}
\end{equation}
Here, $c_i$ is of degree $2i$ and $q_i$ is of degree $4i$.

\item
We discuss the case when $\sG$ is an exceptional group next. 
They can be determined as follows.
We first note that $\bZ$ summands give rise to elements in $H^*(B\sG;\bR)$,
whose structure is known to be dictated by exponents of $\sG$.
The issue is then the torsion part.

For $\sG=F_4,E_{6,7,8}$, the necessary information can be found from the following reasoning. 
Consider the map $f:B\sG\to K(\bZ,4)$ corresponding to the generator of $H^4(B\sG;\bZ)\simeq \bZ$.
The induced map $\pi_d(B\sG)\to \pi_d(K(\bZ,4))$ is an isomorphism when $d<9$
and is surjective when $d=9$.\footnote{%
This is actually an isomorphism unless $\sG=F_4$, for which $\pi_9(BF_4)=\bZ_2$.
Here and elsewhere in the appendix, we need to use the homotopy groups of exceptional groups at a few places.
They can all be found in \cite[Chapter VI, Sec.~7]{MimuraToda}.
}
Therefore $f$ is an $9$-equivalence, and therefore $H_d(B\sG;\bZ)\to H_d(K(\bZ,4);\bZ)$ is an isomorphism when $d\le 8$,\footnote{%
This is a standard fact but here is a quick derivation. Let $f:X\to Y$ be an $n$-equivalence.
Consider the fibration $F(f)\to X\to Y$, where $F(f)$ is the homotopy fiber of $f$.
From the long exact sequence of homotopy groups associated to the fibration,
$\pi_{d<n}(F(f))=0$. Now consider the homology Leray-Serre spectral sequence
$E^2_{p,q}=H_p(Y;H_q(F(f);\bZ))$.
From Hurewicz, the first possibly nonzero $H_{q>0}(F(f))$ is $H_n(F(f))=\pi_n(F(f))$.
Then the spectral sequence in the range $p+q<n$ trivially converges,
and $H_d(X;\bZ)\to H_d(Y;\bZ)$ is an isomorphism at $d<n$.} 
leading to the isomorphism  $H^d(K(\bZ,4);\bZ)\to H^d(BG_i;\bZ)$ when $d\le 8$.
Then, $H^d(K(\bZ,4);\bZ)$  can be found in \cite{BMT13}.
We note that $H_8(B\sG;\bZ)=\bZ\oplus \bZ_3$, so there do appear odd torsions in these cases in higher degrees.

For $\sG=G_2$, it is known that $H^*(BG_2;\bZ)$ is torsion-free for $p\ge 3$
\cite[Chapter VII, Theorem 5.11]{MimuraToda}.
The  structure of $H^*(BG_2;\bZ_2)$ is also known \cite[Chapter VII, Corollary 6.3]{MimuraToda}.
This fixes the entries except at $d=7$, for which we can only conclude $\bZ_{2^k}$ for some $k\ge 1$ at this stage.
Now, consider the map $f:BG_2\to K(\bZ,4)$ as before.
The induced map $\pi_d(BG_2)\to \pi_d(K(\bZ,4))$ is an isomorphism when $d<7$
and is surjective when $d=7$, where $\pi_7(BG_2)=\bZ_3$.
Then $H_d(BG_2;\bZ)\to H_d(K(\bZ,4);\bZ)$ is an isomorphism when $d\le 6$,
and therefore $H^7(BG_2;\bZ)=\bZ_2$.
We note that the differential from $H_8(K(\bZ,4);\bZ)=\bZ\oplus \bZ_3$
to $H_7(F(f);\bZ)=\bZ_3$   in the Leray-Serre spectral sequence for $F(f)\to BG_2\to K(\bZ,4)$
is known to be nontrivial, so that $H_8(BG_2;\bZ)=\bZ$,
in contrast to the case of other exceptional groups, $\sG=F_4, E_{6,7,8}$.

\item
Finally, let us discuss the case  when $\sG=Spin(n)$.
The basic information in this case is (i) that  $H^d(BSpin(n);\bZ)$ is a direct sum of 
copies of $\bZ$ and of $\bZ_2$ \cite{KonoSpin}, and
(ii) that the  structure of $H^*(BSpin(n);\bZ_2)$ is explicitly given in \cite{QuillenBSpin}.
Then, the integral cohomology groups can be found by simply looking at $H^*(BSpin(n);\bZ_2)$
and examining the $Sq^1$ action to see which elements lift to integral cohomology groups.

To actually determine them, we quote the structure of $H^*(BSpin(n);\bZ_2)$ given in \cite{QuillenBSpin}.
Recall that $H^*(BSO(n);\bZ_2)$ is a free polynomial algebra $\bZ_2[w_2,w_3,\ldots,w_n]$.
Then, $H^*(BSpin(n);\bZ_2)$ is obtained by removing the ideal generated by \begin{equation}
w_2\,, \quad
Sq^1 w_2\,, \quad
Sq^2Sq^1 w_2\,,\quad
\ldots\,,\quad
Sq^{2^k} Sq^{2^{k-1}}\cdots Sq^1 w_2\,,\quad
\ldots
\end{equation}
and then by tensoring the algebra  $\bZ_2[\tilde w_{2^h}]$,
where $\tilde w_{2^h}$ is the top Stiefel-Whitney class of the spin representation.
Noting $Sq^1 w_2=w_3$ and $Sq^2 Sq^1 w_2=Sq^2 w_3=w_5+w_2 w_3$,
we find that $H^*(BSpin(n\ge 7);\bZ_2)$ in the range $d\le 8$  is given as follows:
\begin{equation}
\begin{aligned}
H^*(BSpin(7);\bZ_2)&\simeq\bZ_2[w_4, w_6,w_7,\tilde w_8]\, ,\\
H^*(BSpin(8);\bZ_2)&\simeq\bZ_2[w_4, w_6,w_7,w_8,\tilde w_8]\, ,\\
H^*(BSpin(n\ge 9);\bZ_2)&\simeq\bZ_2[w_4, w_6,w_7,w_8]\, .
\end{aligned}
\end{equation}
Here,  $w_i$ are in degree $i$ and come from the Stiefel-Whitney classes of $BSO(n)$,
and $\tilde w_8$ are the 8-th Stiefel-Whitney class of the spin representation.
The action of $Sq^{1,2}$ in this range is given by $Sq^1 w_{2n}=w_{2n+1}$, $Sq^1 w_{2n+1}=0$,
and $Sq^2 w_4=w_6$.
The data given in the table above then follow.

We  note that 
the structure of $H^d(BSpin(n);\bZ)$ is identical to that of $H^d(K(\bZ,4);\bZ)$ when $d\le 7$, 
just as when $\sG$ is an exceptional group.
To see this, we consider the map $f:BSpin(n)\to K(\bZ,4)$ as before.
The induced map $\pi_d(BSpin(n))\to \pi_d(K(\bZ,4))$ is an isomorphism when $d<8$
and is surjective when $d=8$, where $\pi_8(BSpin(n))=\bZ$ or $\bZ^2$ 
depending on whether $n\neq 8$ or $n=8$.
Then $H_d(BSpin(n);\bZ)\to H_d(K(\bZ,4);\bZ)$ is an isomorphism when $d\le 7$,
and therefore $H^d(K(\bZ,4);\bZ)\to H^d(BSpin(n);\bZ)$ is also an isomorphism when $d\le 7$.
\end{itemize}

Our discussions so far also give the structure of $H^d(B\sG;\bZ_2)$ below $d\le 7$ as follows:
\begin{equation}
\begin{aligned}
H^*(BU(1);\bZ_2)&\simeq\bZ[c_1]\, ,\\
H^*(BSU(n\ge 3);\bZ_2)&\simeq\bZ_2[c_2,c_3]\, ,\\
H^*(BSp(n);\bZ_2)&\simeq\bZ_2[q_1]\, ,\\
H^*(B\sG;\bZ_2)&\simeq\bZ_2[w_4,w_6,w_7]\, ,\\
\end{aligned}
\label{z2-coho}
\end{equation}
where $\sG$ in the last line is either $Spin(n\ge 7)$ or one of exceptional groups.
Here, $c_i$ is at degree $2i$, $q_i$ is at degree $4i$,
and $w_i$ is at degree $i$.
Most of the actions of $Sq^1$ and $Sq^2$ vanish due to degree reasons.
$Sq^2 c_3$, $Sq^2 w_6$, $Sq^1 w_7$ also vanish, although
they are degree 8 and we did not give the information on $H^8(B\sG;\bZ_2)$ above.
The nonzero ones in degree $d\le 7$ are \begin{equation}
Sq^2 c_1=c_1^2\,, \quad 
Sq^2 c_2=c_3\,,\quad
Sq^2 w_4=w_6\,,\quad
Sq^1 w_6=w_7\,.
\label{sq-action}
\end{equation}
These all follow from Wu formulas:
\begin{equation}
\begin{aligned}
	Sq^{2i}(c_j) &=\sum_{k=0}^{i} \binom{j-k-1}{i-k} c_{i+j-k}c_{k}\,, \quad (0\leq i\leq j)\,, \\
	Sq^i(w_j) &=\sum_{k=0}^{i} \binom{j-k-1}{i-k} w_{i+j-k}w_{k}\,, \quad (0\leq i\leq j)\,.
\end{aligned}
\end{equation}

\subsection{$H_{d\le 7}(BG;\bZ)$ for product groups $G$ }
We now consider $H_{d}(BG;\bZ)$ where $G=\sG_1\times \cdots \sG_{n-1}$.
This can be computed by a repeated use of the Künneth formula for integral homology \cite[Sec.~3.B]{Hatcher}:
\begin{equation}
H_n(X\times Y;\bZ)\simeq \bigoplus_{p+q=n}H_p(X;H_q(Y;\bZ))\,.
\label{integral-kuenneth}
\end{equation}
We note that the direct sum decomposition on the right hand side is not canonical,
but this is enough for our purposes.

We establish the following properties, used in the next subsection:
\begin{enumerate}
\item $H_d(BG;\bZ)$ in the range $d\le 5$ is torsion-free and nonzero only when $d$ is even.
\item $H_6(BG;\bZ)$ can have torsion, but it is always of the form $\bZ_2^k$.
\item  $H_7(BG;\bZ)=0$.
\end{enumerate}
Let us show these properties by mathematical induction.
First, when $G=\sG$, we can simply check that all three properties  hold
by examining Table~\ref{tab:integ-coho}.

Now assume that we have established these three properties for $G_{n-1}:=\sG_1\times\cdots \times\sG_{n-1}$.
Let us take $G_n:=G_{n-1}\times \sG_{n}$.
We have \begin{equation}
H_d(BG_{n-1}\times B\sG_n;\bZ)
\simeq\bigoplus_{p+q=d} 
H_p(BG_{n-1};H_q(B\sG_n;\bZ))\,.
\end{equation}
Let us consider three properties in turn:
\begin{itemize}
\item
The first property readily follows.
\item
For the second property, possibly nonzero contributions 
on the right hand side is $(p,q)=(0,6)$,
$(2,4)$, $(4,2)$ and $(6,0)$.
In each case, we can either have $\bZ_2^a\oplus \bZ^b$ for some $a$ or $b$.
\item
For the third property, the contributions from $(p,q)=(0,7)$ and $(7,0)$ vanish
by inductive assumptions. 
All other $(p,q)$ are such that either $p\le 5$ is odd or $q\le 5$ is odd 
and therefore the contributions from these vanish.
\end{itemize}
This concludes our proof of the three properties above.

\subsection{Computation of $\Omega^\text{spin}_{7}(BG)$}

The computation from here can be done using either the Atiyah-Hirzebruch spectral sequence
or the  Adams spectral sequence.
The computations in both cases are essentially the same.
Here we opt  for  the Atiyah-Hirzebruch spectral sequence, which is slightly more elementary.

Now,  the Atiyah-Hirzebruch spectral sequence converging to the spin bordism group has the form
\begin{equation}
E^2_{p,q}\simeq H_p(BG;\Omega^\text{spin}_q(pt))\Longrightarrow \Omega^\text{spin}_{p+q}(BG)\, .
\end{equation}
Note that nonzero $\Omega^\text{spin}_{d\le 7}(pt)$
are $\bZ$, $\bZ_2$, $\bZ_2$, $\bZ$ for $d=0,1,2,4$, respectively.
Combining this with the three properties of $H_{d\le 7}(BG;\bZ)$ 
established in the previous subsection,
the $E^2$ page of the spectral sequence has the form given in Table~\ref{tab:AHSS}.

\begin{table}
\[
\begin{array}{c|cccccccccc}
  7 & \cellcolor{wo} 0 &0 \\
  6 & 0 &\cellcolor{wo}  0 & 0\\
  5 & 0 & 0 &\cellcolor{wo} 0  &  0 \\
  4 & \bZ & 0 & * & \cellcolor{wo} 0 & *\\
  3 & 0 & 0 & 0 & 0 & \cellcolor{wo} 0 & 0\\
  2 & \bZ_2 & 0 & * & 0 & \cellcolor{wb}\tikzmark{42}{$*$} & \cellcolor{wo} 0 & * \\
  1 & \bZ_2 & 0 & * & 0 & * & 0 & \cellcolor{wo} \tikzmark{61}{$*$} & * \\
  0 & \bZ & 0 & * & 0 & * & 0 & * & \cellcolor{wo} 0 &\cellcolor{wb} \tikzmark{80}{$*$} \\ 
  \hline
q/p  & 0 & 1 & 2 & 3 & 4 & 5 & 6 & 7 & 8 
  \end{array}\\
\]
\begin{tikzpicture}[overlay,remember picture]
\draw[->] (80)--(61);
\draw[->] (61)--(42);
\end{tikzpicture}
\caption{ The $E^2$ page of the Atiyah-Hirzebruch spectral sequence
to compute $\Omega^\text{spin}_7(BG)$. 
The asterisks stand for possibly nontrivial entries,
$E^2_{p,q}$ with $p+q=7$ are shaded with light orange,
and the $d^2$ differentials we examine are indicated,
together with the source and the target entries shaded with light blue.
\label{tab:AHSS}}
\end{table}
The only possibly nonzero entry for $E^2_{p,q}$ with $p+q=7$ is \begin{equation}
E^2_{6,1}\simeq H_6(BG;\bZ_2)\,.
\end{equation}
We now have differentials \begin{equation}
d^2:  H_6(BG;\bZ_2) \simeq E^2_{6,1} \to E^2_{4,2} \simeq H_4(BG;\bZ_2)
\label{d2-61}
\end{equation} and \begin{equation}
d^2:  H_8(BG;\bZ) \simeq E^2_{8,0} \to E^2_{6,1} \simeq H_6(BG;\bZ_2)\,,
\label{d2-80}
\end{equation}
so that \begin{equation}
E^3_{6,1}\simeq \mathrm{Ker}\ d^2 / \mathrm{Im}\ d^2\,.
\end{equation}
We show below that this is zero. Once this is established, $E^3_{p,q}=0$  for all $p+q=7$,
and we immediately find $\Omega^\text{spin}_{d\le 7}(BG)=0$.

To compute $E^3_{6,1}$, we use the fact that the duals of \eqref{d2-61} and \eqref{d2-80},
\begin{equation}
d_2: H^4(BG;\bZ_2) \to H^6(BG;\bZ_2)
\end{equation} and \begin{equation}
d_2: H^6(BG;\bZ_2) \to H^8(BG;U(1))\,,
\end{equation}
are given by $Sq^2$ and $Sq^2$ post-composed with
$\iota:H^8(BG;\bZ_2)\to H^8(BG;U(1))$, respectively \cite{Teichner,Zhubr}.
Now, this map $\iota$ is an injection when $H_7(BG;\bZ)$ is empty.\footnote{%
To see this, note that $H^8(BG;\bZ_2)=\mathrm{Hom}(H_8(BG;\bZ),\bZ_2)$ 
and $H^8(BG;U(1))=\mathrm{Hom}(H_8(BG;\bZ),U(1))$ 
from the universal coefficient theorem when $H_7(BG;\bZ)=0$.
Writing $H_8(BG;\bZ)=\bZ^k \oplus \bigoplus_i \bZ_{n_i}$ and examining each factor,
one finds that $\iota$ is indeed an injection.
}
Therefore, the dual of $E^3_{6,1}$ is simply \begin{equation}
\frac{\mathrm{Ker}\,Sq^2:H^6(BG;\bZ_2)\to H^8(BG;\bZ_2)}
{\mathrm{Im}\,Sq^2:H^4(BG;\bZ_2)\to H^6(BG;\bZ_2)}\,.
\end{equation}

Let us show that $\mathrm{Ker}\,Sq^2$ is spanned by elements 
\begin{equation}
 c_3=Sq^2 c_2
\end{equation}
of one of $SU(n\ge 3)$, elements
\begin{equation}
w_6=Sq^2 w_4
\end{equation}
 of one of $\sG_i$ which is either $Spin(n)$ or exceptional,
and elements of the form 
\begin{equation}
c_1^2 \tilde c_1+ c_1 \tilde c_1^2= Sq^2(c_1\tilde c_1)\,,
\end{equation}
where $c_1$ and $\tilde c_1$
are $c_1$ of two distinct $U(1)$ factors in $G$.
Once this is established, our claim immediately follows,
since clearly $\mathrm{Ker}\, Sq^2=\mathrm{Im}\, Sq^2$ in this case.

We establish our statement on $\mathrm{Ker}\, Sq^2$ inductively.
When there is no $U(1)$ factor in $G$, 
 $H^6(BG;\bZ_2)$ is spanned by $c_3$ from each factor of $SU(n\ge 3)$
 and $w_6$ from each factor of $Spin(n\ge 7)$ or exceptional groups.
As $Sq^2c_3=0$ and $Sq^2 w_6=0$, both are in $\mathrm{Ker}\, Sq^2$.

Now, we assume that there is a $U(1)$ factor,
so write $G=U(1)\times G'$.
Any  $a\in H^6(BG;\bZ_2)$
is of the form \begin{equation}
a= x_6+ x_4 c_1 + x_2 c_1^2 + x_0 c_1^3 
\end{equation} where $x_d \in H^d(BG';\bZ_2)$. We have \begin{equation}
Sq^2 a = (Sq^2 x_6)+ (Sq^2 x_4) c_1  + (x_4 + Sq^2 x_2)c_1^2+ x_0 c_1^4\,.
\end{equation}
Setting $Sq^2a=0$, we find that \begin{equation}
Sq^2 x_6=0, \quad Sq^2 x_4=0, \quad x_4=Sq^2 x_2, \quad x_0=0\,.
\end{equation}
The part $Sq^2 x_6=0$ is taken care of by the inductive hypothesis.
Then $x_2$ is a linear combination of $c_1$ of other $U(1)$ factors in $G$.
Therefore, $a$ is indeed spanned by elements of the form \begin{equation}
(Sq^2 \tilde c_1) c_1 + \tilde c_1 c_1^2 = c_1^2 \tilde c_1+ c_1\tilde c_1^2\,.
\end{equation}
This is what we wanted to show.

\end{appendix}

\newpage

\providecommand{\href}[2]{#2}\begingroup\raggedright\endgroup


\begin{thebibliography}{10}

\bibitem{Monnier:2018nfs}
S.~Monnier and G.~W. Moore, \emph{{Remarks on the Green\textendash{}Schwarz Terms of Six-Dimensional Supergravity Theories}}, \href{https://doi.org/10.1007/s00220-019-03341-7}{\emph{Commun. Math. Phys.} {\bfseries 372} (2019) 963} [\href{https://arxiv.org/abs/1808.01334}{{\ttfamily 1808.01334}}].

\bibitem{Vafa:2005ui}
C.~Vafa, \emph{{The String landscape and the swampland}},  \href{https://arxiv.org/abs/hep-th/0509212}{{\ttfamily hep-th/0509212}}.

\bibitem{Avramis:2005hc}
S.~D. Avramis and A.~Kehagias, \emph{{A Systematic search for anomaly-free supergravities in six dimensions}}, \href{https://doi.org/10.1088/1126-6708/2005/10/052}{\emph{JHEP} {\bfseries 10} (2005) 052} [\href{https://arxiv.org/abs/hep-th/0508172}{{\ttfamily hep-th/0508172}}].

\bibitem{Randjbar-Daemi:1985tdc}
S.~Randjbar-Daemi, A.~Salam, E.~Sezgin and J.~A. Strathdee, \emph{{An Anomaly Free Model in Six-Dimensions}}, \href{https://doi.org/10.1016/0370-2693(85)91653-3}{\emph{Phys. Lett. B} {\bfseries 151} (1985) 351}.

\bibitem{Avramis:2005qt}
S.~D. Avramis, A.~Kehagias and S.~Randjbar-Daemi, \emph{{A New anomaly-free gauged supergravity in six dimensions}}, \href{https://doi.org/10.1088/1126-6708/2005/05/057}{\emph{JHEP} {\bfseries 05} (2005) 057} [\href{https://arxiv.org/abs/hep-th/0504033}{{\ttfamily hep-th/0504033}}].

\bibitem{Becker:2023zyb}
K.~Becker, A.~Kehagias, E.~Sezgin, D.~Tennyson and A.~Violaris, \emph{{New anomaly free supergravities in six dimensions}},  \href{https://arxiv.org/abs/2311.03337}{{\ttfamily 2311.03337}}.

\bibitem{Kumar:2010ru}
V.~Kumar, D.~R. Morrison and W.~Taylor, \emph{{Global aspects of the space of 6D N = 1 supergravities}}, \href{https://doi.org/10.1007/JHEP11(2010)118}{\emph{JHEP} {\bfseries 11} (2010) 118} [\href{https://arxiv.org/abs/1008.1062}{{\ttfamily 1008.1062}}].

\bibitem{Seiberg:2011dr}
N.~Seiberg and W.~Taylor, \emph{{Charge Lattices and Consistency of 6D Supergravity}}, \href{https://doi.org/10.1007/JHEP06(2011)001}{\emph{JHEP} {\bfseries 06} (2011) 001} [\href{https://arxiv.org/abs/1103.0019}{{\ttfamily 1103.0019}}].

\bibitem{Monnier:2017oqd}
S.~Monnier, G.~W. Moore and D.~S. Park, \emph{{Quantization of anomaly coefficients in 6D $\mathcal{N}=(1,0)$ supergravity}}, \href{https://doi.org/10.1007/JHEP02(2018)020}{\emph{JHEP} {\bfseries 02} (2018) 020} [\href{https://arxiv.org/abs/1711.04777}{{\ttfamily 1711.04777}}].

\bibitem{Monnier:2018cfa}
S.~Monnier and G.~W. Moore, \emph{{A Brief Summary Of Global Anomaly Cancellation In Six-Dimensional Supergravity}},  \href{https://arxiv.org/abs/1808.01335}{{\ttfamily 1808.01335}}.

\bibitem{Pang:2020rir}
Y.~Pang and E.~Sezgin, \emph{{On the consistency of a class of R-symmetry gauged 6$D$ $\mathcal{N}$=(1,0) supergravities}}, \href{https://doi.org/10.1098/rspa.2020.0115}{\emph{Proc. Roy. Soc. Lond. A} {\bfseries 476} (2020) 20200115} [\href{https://arxiv.org/abs/2002.04619}{{\ttfamily 2002.04619}}].

\bibitem{Alvarez-Gaume:1984zlq}
L.~Alvarez-Gaume and P.~H. Ginsparg, \emph{{The Structure of Gauge and Gravitational Anomalies}}, \href{https://doi.org/10.1016/0003-4916(85)90087-9}{\emph{Annals Phys.} {\bfseries 161} (1985) 423}.

\bibitem{Okubo:1981td}
S.~Okubo, \emph{{Modified Fourth Order Casimir Invariants and Indices for Simple Lie Algebras}}, \href{https://doi.org/10.1063/1.525212}{\emph{J. Math. Phys.} {\bfseries 23} (1982) 8}.

\bibitem{McKayPateraBook}
W.~G. McKay and J.~Patera, \emph{Tables of Dimensions, Indices, and Branching Rules for Representations of Simple Lie Algebras}. Marcel Dekker Inc., New York, 1981.

\bibitem{Feger:2019tvk}
R.~Feger, T.~W. Kephart and R.~J. Saskowski, \emph{{LieART 2.0 {\textendash} A Mathematica application for Lie Algebras and Representation Theory}}, \href{https://doi.org/10.1016/j.cpc.2020.107490}{\emph{Comput. Phys. Commun.} {\bfseries 257} (2020) 107490} [\href{https://arxiv.org/abs/1912.10969}{{\ttfamily 1912.10969}}].

\bibitem{Nishino:1986Dc}
H.~Nishino and E.~Sezgin, \emph{{The Complete $N=2$, $d=6$ Supergravity With Matter and {Yang-Mills} Couplings}}, \href{https://doi.org/10.1016/0550-3213(86)90218-X}{\emph{Nucl. Phys. B} {\bfseries 278} (1986) 353}.

\bibitem{Ferrara:1997gh}
S.~Ferrara, F.~Riccioni and A.~Sagnotti, \emph{{Tensor and vector multiplets in six-dimensional supergravity}}, \href{https://doi.org/10.1016/S0550-3213(97)00837-7}{\emph{Nucl. Phys. B} {\bfseries 519} (1998) 115} [\href{https://arxiv.org/abs/hep-th/9711059}{{\ttfamily hep-th/9711059}}].

\bibitem{Riccioni:1998th}
F.~Riccioni and A.~Sagnotti, \emph{{Consistent and covariant anomalies in six-dimensional supergravity}}, \href{https://doi.org/10.1016/S0370-2693(98)00846-6}{\emph{Phys. Lett. B} {\bfseries 436} (1998) 298} [\href{https://arxiv.org/abs/hep-th/9806129}{{\ttfamily hep-th/9806129}}].

\bibitem{Bossard:2024ffp}
G.~Bossard, A.~Kleinschmidt and E.~Sezgin, \emph{{Higher derivative couplings with multi-tensor multiplets in 6D supergravity, action and anomalies}},  \href{https://arxiv.org/abs/2412.05365}{{\ttfamily 2412.05365}}.

\bibitem{Riccioni:2001bg}
F.~Riccioni, \emph{{All couplings of minimal six-dimensional supergravity}}, \href{https://doi.org/10.1016/S0550-3213(01)00199-7}{\emph{Nucl. Phys. B} {\bfseries 605} (2001) 245} [\href{https://arxiv.org/abs/hep-th/0101074}{{\ttfamily hep-th/0101074}}].

\bibitem{Cheung:2016wjt}
C.~Cheung and G.~N. Remmen, \emph{{Positivity of Curvature-Squared Corrections in Gravity}}, \href{https://doi.org/10.1103/PhysRevLett.118.051601}{\emph{Phys. Rev. Lett.} {\bfseries 118} (2017) 051601} [\href{https://arxiv.org/abs/1608.02942}{{\ttfamily 1608.02942}}].

\bibitem{Hamada:2018dde}
Y.~Hamada, T.~Noumi and G.~Shiu, \emph{{Weak Gravity Conjecture from Unitarity and Causality}}, \href{https://doi.org/10.1103/PhysRevLett.123.051601}{\emph{Phys. Rev. Lett.} {\bfseries 123} (2019) 051601} [\href{https://arxiv.org/abs/1810.03637}{{\ttfamily 1810.03637}}].

\bibitem{Caron-Huot:2021rmr}
S.~Caron-Huot, D.~Mazac, L.~Rastelli and D.~Simmons-Duffin, \emph{{Sharp boundaries for the swampland}}, \href{https://doi.org/10.1007/JHEP07(2021)110}{\emph{JHEP} {\bfseries 07} (2021) 110} [\href{https://arxiv.org/abs/2102.08951}{{\ttfamily 2102.08951}}].

\bibitem{Bobev:2021qxx}
N.~Bobev, K.~Hristov and V.~Reys, \emph{{AdS$_{5}$ holography and higher-derivative supergravity}}, \href{https://doi.org/10.1007/JHEP04(2022)088}{\emph{JHEP} {\bfseries 04} (2022) 088} [\href{https://arxiv.org/abs/2112.06961}{{\ttfamily 2112.06961}}].

\bibitem{Suzuki:2005vu}
R.~Suzuki and Y.~Tachikawa, \emph{{More anomaly-free models of six-dimensional gauged supergravity}}, \href{https://doi.org/10.1063/1.2209767}{\emph{J. Math. Phys.} {\bfseries 47} (2006) 062302} [\href{https://arxiv.org/abs/hep-th/0512019}{{\ttfamily hep-th/0512019}}].

\bibitem{Randjbar-Daemi:2004bjl}
S.~Randjbar-Daemi and E.~Sezgin, \emph{{Scalar potential and dyonic strings in 6-D gauged supergravity}}, \href{https://doi.org/10.1016/j.nuclphysb.2004.05.023}{\emph{Nucl. Phys. B} {\bfseries 692} (2004) 346} [\href{https://arxiv.org/abs/hep-th/0402217}{{\ttfamily hep-th/0402217}}].

\bibitem{Polchinski:2003bq}
J.~Polchinski, \emph{{Monopoles, duality, and string theory}}, \href{https://doi.org/10.1142/S0217751X0401866X}{\emph{Int. J. Mod. Phys. A} {\bfseries 19S1} (2004) 145} [\href{https://arxiv.org/abs/hep-th/0304042}{{\ttfamily hep-th/0304042}}].

\bibitem{Banks:2010zn}
T.~Banks and N.~Seiberg, \emph{{Symmetries and Strings in Field Theory and Gravity}}, \href{https://doi.org/10.1103/PhysRevD.83.084019}{\emph{Phys. Rev. D} {\bfseries 83} (2011) 084019} [\href{https://arxiv.org/abs/1011.5120}{{\ttfamily 1011.5120}}].

\bibitem{Hellerman:2010fv}
S.~Hellerman and E.~Sharpe, \emph{{Sums over topological sectors and quantization of Fayet-Iliopoulos parameters}}, \href{https://doi.org/10.4310/ATMP.2011.v15.n4.a7}{\emph{Adv. Theor. Math. Phys.} {\bfseries 15} (2011) 1141} [\href{https://arxiv.org/abs/1012.5999}{{\ttfamily 1012.5999}}].

\bibitem{Deser:1997se}
S.~Deser, A.~Gomberoff, M.~Henneaux and C.~Teitelboim, \emph{{P-brane dyons and electric magnetic duality}}, \href{https://doi.org/10.1016/S0550-3213(98)00179-5}{\emph{Nucl. Phys. B} {\bfseries 520} (1998) 179} [\href{https://arxiv.org/abs/hep-th/9712189}{{\ttfamily hep-th/9712189}}].

\bibitem{Tosa:1989qm}
Y.~Tosa, \emph{{Global gauge anomalies for theories with the Green-Schwarz local anomaly cancellation mechanism}}, \href{https://doi.org/10.1103/PhysRevD.40.1934}{\emph{Phys. Rev. D} {\bfseries 40} (1989) 1934}.

\bibitem{Bershadsky:1997sb}
M.~Bershadsky and C.~Vafa, \emph{{Global anomalies and geometric engineering of critical theories in six-dimensions}},  \href{https://arxiv.org/abs/hep-th/9703167}{{\ttfamily hep-th/9703167}}.

\bibitem{Lee:2020ewl}
Y.~Lee and Y.~Tachikawa, \emph{{Some comments on 6D global gauge anomalies}}, \href{https://doi.org/10.1093/ptep/ptab015}{\emph{PTEP} {\bfseries 2021} (2021) 08B103} [\href{https://arxiv.org/abs/2012.11622}{{\ttfamily 2012.11622}}].

\bibitem{Davighi:2020kok}
J.~Davighi and N.~Lohitsiri, \emph{{Omega vs. pi, and 6d anomaly cancellation}}, \href{https://doi.org/10.1007/JHEP05(2021)267}{\emph{JHEP} {\bfseries 05} (2021) 267} [\href{https://arxiv.org/abs/2012.11693}{{\ttfamily 2012.11693}}].

\bibitem{Guo:2025mlb}
X.~Guo, Y.~Pang and E.~Sezgin, \emph{{4D de Sitter from 6D gauged supergravity with Green-Schwarz counterterm}},  \href{https://arxiv.org/abs/2510.11794}{{\ttfamily 2510.11794}}.

\bibitem{Beaudry:2018ifm}
A.~Beaudry and J.~A. Campbell, \emph{{A Guide for Computing Stable Homotopy Groups}},  \href{https://arxiv.org/abs/1801.07530}{{\ttfamily 1801.07530}}.

\bibitem{Lee:2022spd}
Y.~Lee and K.~Yonekura, \emph{{Global anomalies in 8d supergravity}}, \href{https://doi.org/10.1007/JHEP07(2022)125}{\emph{JHEP} {\bfseries 07} (2022) 125} [\href{https://arxiv.org/abs/2203.12631}{{\ttfamily 2203.12631}}].

\bibitem{MimuraToda}
M.~Mimura and H.~Toda, \emph{Topology of {L}ie groups. {I}, {II}}, vol.~91 of \emph{Translations of Mathematical Monographs}. American Mathematical Society, Providence, RI, 1991.

\bibitem{BMT13}
L.~Breen, R.~Mikhailov and A.~Touz\'{e}, \emph{Derived functors of the divided power functors}, \href{https://doi.org/10.2140/gt.2016.20.257}{\emph{Geom. Topol.} {\bfseries 20} (2016) 257} [\href{https://arxiv.org/abs/1312.5676}{{\ttfamily 1312.5676}}].

\bibitem{KonoSpin}
A.~Kono, \emph{{On the integral cohomology of $BSpin(n)$}}, \href{https://doi.org/10.1215/kjm/1250520870}{\emph{J. Math. Kyoto Univ.} {\bfseries 26} (1986) 333}.

\bibitem{QuillenBSpin}
D.~Quillen, \emph{The {${\rm mod}$} {$2$} cohomology rings of extra-special {$2$}-groups and the spinor groups}, \href{https://doi.org/10.1007/BF01350050}{\emph{Math. Ann.} {\bfseries 194} (1971) 197}.

\bibitem{Hatcher}
A.~Hatcher, \emph{Algebraic Topology}. Cambridge University Press, 2002.

\bibitem{Teichner}
P.~Teichner, \emph{On the signature of four-manifolds with universal covering spin}, \href{https://doi.org/10.1007/BF01444915}{\emph{Math. Ann.} {\bfseries 295} (1993) 745}.

\bibitem{Zhubr}
A.~V. Zhubr, \emph{Spin bordism of oriented manifolds and the {H}auptvermutung for 6-manifolds}, \href{https://doi.org/10.1090/trans2/202/19}{\emph{Amer. Math. Soc. Transl. Ser. 2} {\bfseries 202} (2001) 263}.

\bibitem{Becker:2026etl}
K.~Becker and Q.~You,
{\it Anomaly-Free Spectra, Unimodular Lattices and 6D R-Symmetry Gauged Supergravity,}
[arXiv:2605.17643 [hep-th]].

\end{thebibliography}

\end{document}